\documentclass{aa}  
\usepackage{graphicx}
\usepackage{comment}
\graphicspath{{./figures/}}
\usepackage{txfonts}
\usepackage[dvipsnames]{xcolor}
\usepackage{soul}
\usepackage{natbib,twoopt}
\usepackage[breaklinks=true]{hyperref} %% to avoid \citeads line fills
\bibpunct{(}{)}{;}{a}{}{,}             %% natbib format for A&A and ApJ
\makeatletter
  \newcommandtwoopt{\citeads}[3][][]{\href{http://adsabs.harvard.edu/abs/#3}%
    {\def\hyper@linkstart##1##2{}%
     \let\hyper@linkend\@empty\citealp[#1][#2]{#3}}}
  \newcommandtwoopt{\citepads}[3][][]{\href{http://adsabs.harvard.edu/abs/#3}%
    {\def\hyper@linkstart##1##2{}%
     \let\hyper@linkend\@empty\citep[#1][#2]{#3}}}
  \newcommandtwoopt{\citetads}[3][][]{\href{http://adsabs.harvard.edu/abs/#3}%
    {\def\hyper@linkstart##1##2{}%
     \let\hyper@linkend\@empty\citet[#1][#2]{#3}}}
  \newcommandtwoopt{\citeyearads}[3][][]%
    {\href{http://adsabs.harvard.edu/abs/#3}
    {\def\hyper@linkstart##1##2{}%
     \let\hyper@linkend\@empty\citeyear[#1][#2]{#3}}}
\makeatother
\usepackage[]{hyperref}

\begin{document}

   \title{Modeling the Remnants of Core-collapse Supernovae from Luminous Blue Variable stars}

   \author{S. Ustamujic
          \inst{1}
          \and
          S. Orlando
          \inst{1}
          \and
          M. Miceli
          \inst{2,1}
          \and
          F. Bocchino
          \inst{1}
          \and
          M. Limongi
          \inst{3,4,5}
          \and
          A. Chieffi
          \inst{5,6,7}
          \and
          C. Trigilio
          \inst{8}
          \and
          G. Umana
          \inst{8}
          \and \\
          F. Bufano
          \inst{8}
          \and
          A. Ingallinera
          \inst{8}
          \and
          G. Peres
          \inst{2,1}}

   \institute{INAF-Osservatorio Astronomico di Palermo, Piazza del Parlamento 1,
             90134 Palermo, Italy\\
              \email{sabina.ustamujic@inaf.it}
         \and
             Dipartimento di Fisica e Chimica E. Segr\`e, Universit\`a di Palermo, 
             Via Archirafi 36, 90123 Palermo, Italy
         \and     
             INAF-Osservatorio Astronomico di Roma, Via Frascati 33, I-00040, 
             Monteporzio Catone, Italy
         \and
             Kavli Institute for the Physics and Mathematics of the Universe, 
             Todai Institutes for Advanced Study, University of Tokyo, Kashiwa, 
             277-8583 (Kavli IPMU, WPI), Japan
         \and
             INFN. Sezione di Perugia, via A. Pascoli s/n, I-06125 Perugia, Italy
         \and
             INAF-Istituto di Astrofisica e Planetologia Spaziali, 
             Via Fosso del Cavaliere 100, I-00133, Roma, Italy
         \and
             Monash Centre for Astrophysics (MoCA), School of Mathematical 
             Sciences, Monash University, VIC 3800, Australia
         \and     
             INAF-Osservatorio Astrofisico di Catania, Via Santa Sofia 78, 95123 Catania, Italy
             }

   \date{Received 16 June 2021 / Accepted 3 August 2021}
 
  \abstract
  % context heading (optional)
   {Luminous Blue Variable stars (LBVs) are massive evolved stars that 
   suffer sporadic and violent mass-loss events. They have been 
   proposed as the progenitors of some core-collapse supernovae 
   (SNe), but this idea is still debated due to the lack of direct 
   evidence. Since SNRs can carry in their morphology the fingerprints of the progenitor stars as well as of the inhomogeneous circumstellar medium (CSM) sculpted by the progenitors, the study of SNRs from LBVs could help to place 
   core-collapse SNe in context with the evolution of massive stars.}  
  % aims heading (mandatory)
   {We investigate the physical, chemical and morphological 
   properties of the remnants of SNe originating from LBVs, in order to search for signatures, revealing the nature of the progenitors, in the ejecta distribution and morphology of the remnants.}
  % methods heading (mandatory)
   {As a template of LBVs, we considered the actual LBV candidate Gal 026.47+0.02. We selected 
   a grid of models, which describe the evolution of a massive star 
   with properties consistent with those of Gal 026.47+0.02 and its final fate as core-collapse SN. We developed a three-dimensional (3D) 
   hydrodynamic (HD) model that follows the post-explosion evolution 
   of the ejecta from the breakout of the shock wave at the stellar surface to the 
   interaction of the SNR with a CSM characterized by two dense nested toroidal shells, 
   parametrized in agreement with multi-wavelength observations of Gal 026.47+0.02.}
  % results heading (mandatory)
   {Our models show a strong interaction of the blast wave with the CSM 
   which determines an important slowdown of the expansion of the ejecta 
   in the equatorial plane where the two shells lay, determining a high degree 
   of asymmetry in the remnant. After $\approx 10000$~years of evolution 
   the ejecta show an elongated shape forming a broad jet-like structure caused 
   by the interaction with the shells and oriented along the axis of the 
   toroidal shells. Models with high explosion energy show 
   Fe-rich internal ejecta distributions surrounded by an elongated 
   Si-rich structure with a more diffuse O-rich ejecta all around. 
   Models with low explosion energy instead show a more homogeneous distribution of 
   chemical elements with a very low presence of Fe-group elements.}
  % conclusions heading (optional), leave it empty if necessary 
   {The geometry and density distribution of the CSM where a LBV star goes SN are fundamental 
   in determining the properties of the resulting SNR. For all the 
   LBV-like progenitors explored here, we found that the remnants show a common morphology, namely 
   elongated ejecta with an internal jet-like structure, which reflects the inhomogeneous and dense pre-SN CSM surrounding the star.}


   \maketitle
%
%--------------------------------------------------------------------
\section{Introduction}
\label{Sec:intro}

  Luminous Blue Variable stars (LBVs) are massive evolved unstable stars that strongly interact with the circumstellar medium (CSM), showing dramatic variations in both their spectra and brightness \citep{hum94,hum99}.
  This class includes stars displaying variability on different timescales and
  intensities \citep{dek96}. The most typical LBVs are the S Doradus variables,
  which are characterized by $\approx \,0.5-2$~mag quasi-periodic variations on 
  a timescale of years to decades. 
  Giant eruptions instead are much less common than the variability observed in S Doradus variables and determine greater brightness variations ($\geq 1-2$~mag) associated with an episode of high mass-loss ($\sim10\,M_\odot$).
  Examples of these giant eruptions are those occurred in the case of 
  $\eta$-Carinae \citep[see][]{dav97,smi18}.
  Although we still do not fully understand the physical
  mechanism that drives LBV variability, some progress has been made in the last years. Recently \cite{gra21} have developed a model that reproduces the typical observational phenomenology of the S Doradus variability. According to their model, the instability responsible for the observed variability can be triggered when some physical conditions are met, involving inflated envelopes in proximity to the Eddington limit, a temperature range that does not lead to accelerating outflows, and a mass-loss rate that increases with decreasing temperature \citep[see][for more details]{gra21}.
  The causes of the sporadic and violent mass-loss events, however, are still poorly 
  understood and a physical driving mechanism has not been clearly identified yet \citep{smi18}.
  
  According to the traditional stellar evolution theory, LBVs are presumed to represent a brief transitional 
  phase in the evolution of the most massive stars, between the main-sequence 
  O-type stars and the Wolf–Rayet (WR) stars, but in the last decade the 
  theoretical understanding of massive stellar evolution has been radically
  revised \citep[see][]{smi17}. In particular, in a number of recent works,
  LBVs have been proposed as the progenitors of some core-collapse
  supernovae (SNe), particularly Type IIn \citep[e.g.][]{mil10,smi11,fox11,tad13,pas18,tad20} and Type IIb \citep[e.g.][]{gro13, mor13, pre20} SNe.
  In many cases, the link with SNe stems from the fact that, in this class of SNe, the blast wave appears to expand in a high-density medium, which has been interpreted as 
  resulting from a wind with velocities and mass-loss rates consistent
  with those expected for massive LBVs. 
  However discriminating between some stellar transients and the SN events 
  can be a tricky issue in some cases, considering the amount of energy released
  during some stellar outbursts \citep[see][]{pas19_nat}.
  Indeed during giant eruptive episodes, LBVs can reach total luminosities 
  comparable to that of a SN, mimicking in some cases the behaviour of real SNe IIn; these events (that have nothing to do with SNe) are known as SN impostors \citep{van00,sma09,van12,tar15}.
  
  Despite the similarities between the characteristics of some SNe (namely those showing strong shocks resulting from the interaction between their ejecta and the pre-existing dense CSM) and those of LBVs (more specifically, the giant eruption of mass and the highly structured and dense CSM), a firm evidence that LBVs are direct core-collapse 
  SNe progenitors is still lacking \citep{dwa11,smi17,van17,nyh20}. In fact, a large population of sources closely related to the LBV stage of massive 
  stellar evolution have been identified \citep{wac10,gva10,gva15,smi19}, 
  but their connection with some Type IIn/IIb SNe is not clear. 
  On the other hand, the information about the environments in which the 
  stars explode may establish some constraints on the evolutionary 
  phases of the progenitors \citep{fox11,and12,nyh20}. However, while observations show that some SNe seem to have had LBVs as progenitors, 
  there are only a couple of theoretical models that support LBVs exploding 
  as SNe in the framework of single stellar evolution \citep{gro13,mor13}.
  
  Another interesting clue to be explored is the possible identification
  of signatures of LBV progenitors in the ejecta distribution and morphology of 
  supernova remnants (SNRs).
  In fact, the morphology and the distribution of material observed in SNRs 
  is expected to reflect the interaction of the SN blast wave with the 
  ambient environment \citep[e.g.][]{ust21}, the physical processes associated with the SN explosion and the nature of the progenitor star. 
  Being able to disentangle the different effects is a challenging 
  puzzle to explore and solve \citep[e.g.,][]{orl15,orl16,won17,fer19,tut20,orl19,orl20,orl21,gab21,fer21,jac21}.
  Recently, \cite{chi21} proposed that the two protrusions that are projected as two ``ears'' in the morphology of some SNRs could be formed by the interaction of the remnant with the CSM, considering a LBV or a red/yellow supergiant as the most 
  likely progenitors in case of a core collapse SN event. Other authors 
  instead support the scenario of the jet-driven core-collapse SN 
  mechanisms to explain these elongated features \citep[see][]{gri17,bea18}.
  
  In this work, we investigate the physical, chemical and morphological 
  properties of SNRs of stars exploded as SNe at (or soon after) the LBV 
  evolutionary stage. Generally LBVs are surrounded by extended 
  circumstellar envelopes that show a wide variety of characteristics reflecting the mass-loss history of the variable stars.
  For our study, we adopted as a template the LBV candidate Gal 026.47+0.02 \citep[hereafter G26; see][]{cla03,cla05}, which exhibits one of the highest 
  observed mass losses from the central object. G26 is located in a very 
  massive nebula that has been extensively studied and, for which, an accurate description of its structure and density distribution has been derived from the analysis of the observations \citep[see][]{par12,uma12}.
  The extreme characteristics of the environment associated with G26 
  \citep[see][]{uma12} fit quite well with the requirements for the 
  progenitor of Type IIn SN 2010jl, according to the model by \cite{and11}. 
  This may indicate that G26 could be a precursor of a very bright Type IIn SN. 
  The central star shows a luminosity 
  $\log(L/L_\odot)=6$, a temperature $T=17000$~K and mass ejection 
  rate $\dot M=9\times 10^{-5} M_\odot$~$yr^{-1}$ \citep{cla03,wac10}. 
  Assuming a distance of 6.5~kpc \citep{cla03}, the nebula G26 consists 
  of $\approx 20\,M_\odot$ of ionized gas distributed in two nested tori (or shells)
  around a common axis which have been interpreted as the observational 
  evidence of past episodic mass loss events \citep{uma12}. 
  Its temperature and luminosity place G26 in a region of the HR diagram which is populated by quite typical LBV stars (see Figs.~1~and~3 in \citealt{smi17}). The high total mass of material detected in G26 could be an indication of a few giant energetic eruptions occurred in the past or a number of moderate mass loss events \citep{uma12}. Thus, the distribution of material observed in the circumstellar environment reflects the mass-loss history of the star and, therefore, is peculiar of G26. According to evolutionary models, G26 evolved from a star with an initial 
  mass between 60 and 80 $M_\odot$ (see Fig.~14 in \citealt{lim18}; 
  see also \citealt{smi17}).
  
  For our purposes, we selected a grid of massive stars, in agreement with the characteristics of G26, which explode as core-collapse SNe from the stellar models described in \cite{lim18}. Then, we performed three-dimensional (3D) hydrodynamic (HD) simulations that follow the post-explosion evolution of the SN from the breakout of the shock wave at the surface of the LBV progenitor to the interaction of the SNR with the cirsumstellar environment. We explored a grid of eight models differing as for the progenitor star characteristics and the explosion energy, and considering the same circumstellar environment described in \cite{uma12}.

  The paper is organized as follows. In Sect. 2 we describe the
  model and the numerical setup; in Sect. 3 we discuss the results; and 
  in Sect. 4 we draw our conclusions.

%--------------------------------------------------------------------
\section{Hydrodynamic model}
\label{Sec:model}

  The model describes the post-explosion evolution of a core-collapse SN
  from the breakout of the shock wave at the stellar surface (occurring 
  a few minutes after the SN event) to the interaction of the blast wave 
  and ejecta caused by the explosion with the circumstellar environment.
  We followed the evolution for $t\approx 10000$~yr by numerically 
  solving the full time-dependent HD equations in a 3D Cartesian coordinate 
  system $(x, y, z)$.
  The HD equations were solved in the conservative form
  
   \begin{equation}
      \frac{\partial \rho}{\partial t} 
      + \nabla \cdot (\rho \boldsymbol{u}) = 0,
   \end{equation}
   \begin{equation}
      \frac{\partial (\rho \boldsymbol{u}) }{\partial t} + \nabla \cdot 
      (\rho\boldsymbol{u}\boldsymbol{u}) 
      + \nabla P = 0,
   \end{equation}
   \begin{equation}
      \frac{\partial (\rho E) }{\partial t}
      + \nabla \cdot [\boldsymbol{u} (\rho E+P)] = 0,
   \label{eq.energy}
   \end{equation}
   where $E = \epsilon + u^2/2$ is the total gas energy (internal energy 
   $\epsilon$, and kinetic energy) per unit mass, $t$ is the time, 
   $\rho = \mu m_{\mathrm{H}} n$ is the mass density, 
   $\mu$ is the mean atomic mass ($\mu$ for the ejecta considers their isotopic composition, whereas $\mu = 1.29$ for the CSM, assuming cosmic abundances), 
   $m_{\mathrm{H}}$ is the mass of the hydrogen atom, 
   $n$ is the total number density, $\boldsymbol{u}$ is 
   the gas velocity, and $T$ is the temperature. We used the ideal 
   gas law, $P = (\gamma-1)\rho\epsilon$, where $\gamma=5/3$ is the adiabatic 
   index.

   The calculations were performed using PLUTO \citep{mig07}, a modular 
   Godunov-type code for astrophysical plasmas. The code provides a 
   multiphysics, multialgorithm modular environment particularly oriented 
   towards the treatment of astrophysical high Mach number flows in 
   multiple spatial dimensions. The code was designed to make efficient 
   use of massive parallel computers using the message-passing interface
   (MPI) library for interprocessor communications. The HD equations are
   solved using the HD module available in PLUTO; the integration is 
   performed using the original Piecewise Parabolic Method (PPM)
   reconstruction by \citet[see also \citealt{mil02}]{col84} with a Roe 
   Riemann solver. The adopted scheme is particularly appropriate for 
   describing the shocks formed during the interaction of the remnant 
   with the surrounding inhomogeneous medium, as in our case.
   A monotonous central difference limiter (the least diffusive limiter
   available in PLUTO) for the primitive variables is used.
   The code was extended by additional computational modules to evaluate
   the deviations from equilibrium of ionization of the most abundant ions
   (through the computation of the maximum ionization age in each cell 
   of the spatial domain as described in \citealt{orl15}), and the deviations 
   from temperature-equilibration between electrons and ions. 
   For the latter, we included the almost instantaneous heating of 
   electrons at shock fronts up to $kT \sim 0.3$~keV by lower 
   hybrid waves \citep[see][]{gha07}, and the effects of Coulomb collisions 
   for the calculation of ion and electron temperatures in the post-shock 
   plasma \citep[see][for further details]{orl15}.

\subsection{Initial and boundary conditions}
\label{Sec:num}

\begin{figure*}
    \includegraphics*[width=0.5\hsize]{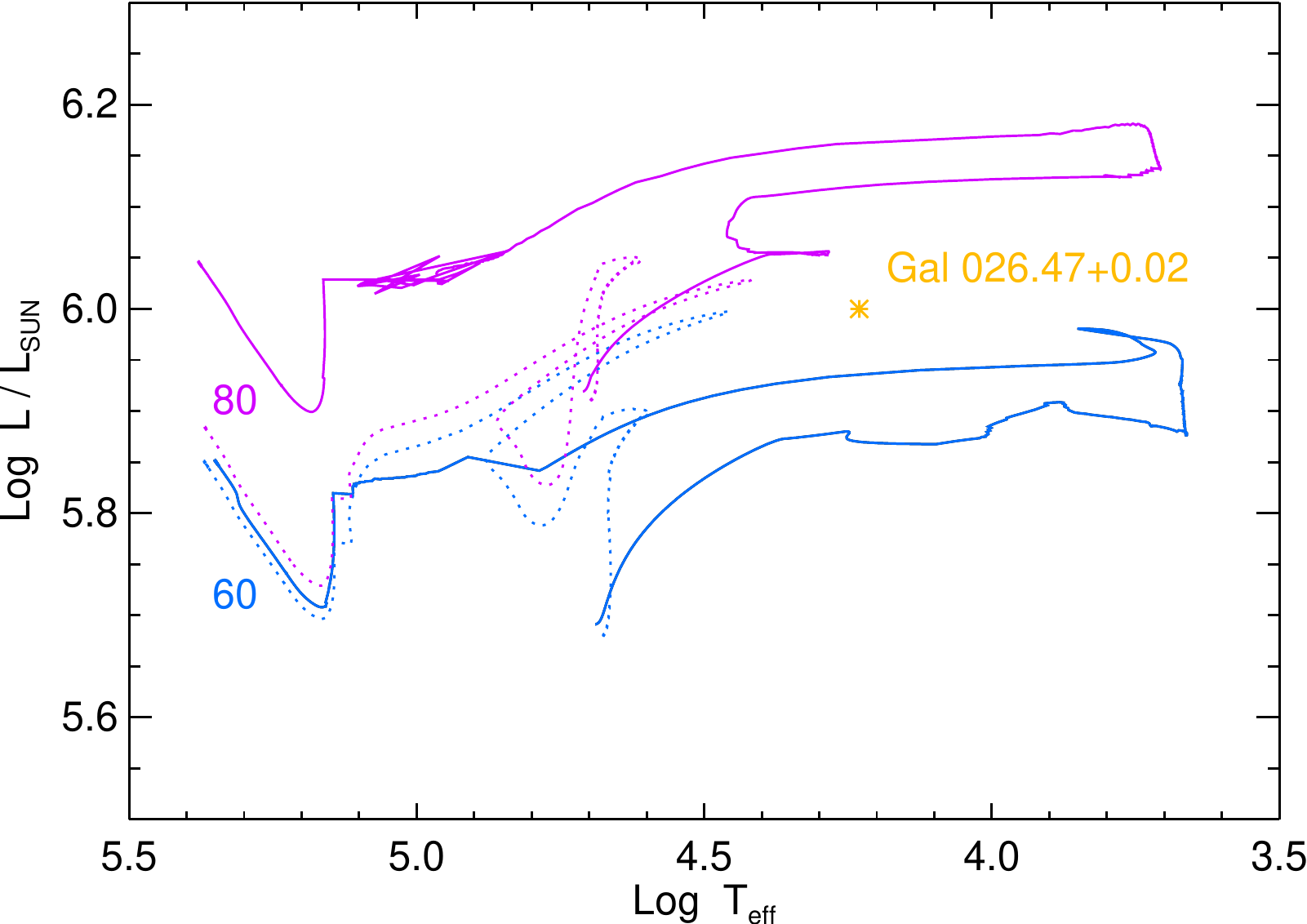}
    \includegraphics*[width=0.5\hsize]{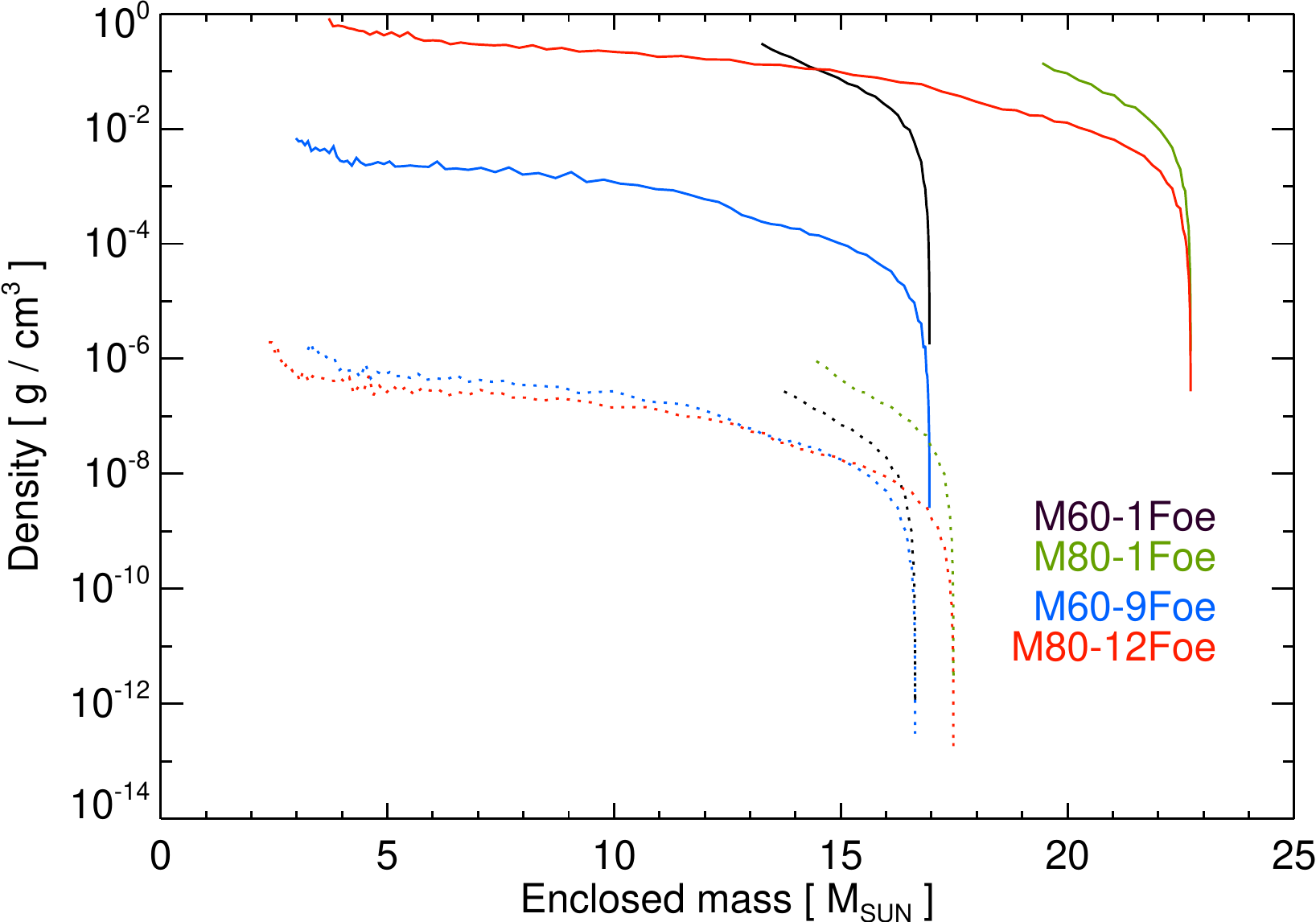}
    \includegraphics*[width=0.5\hsize]{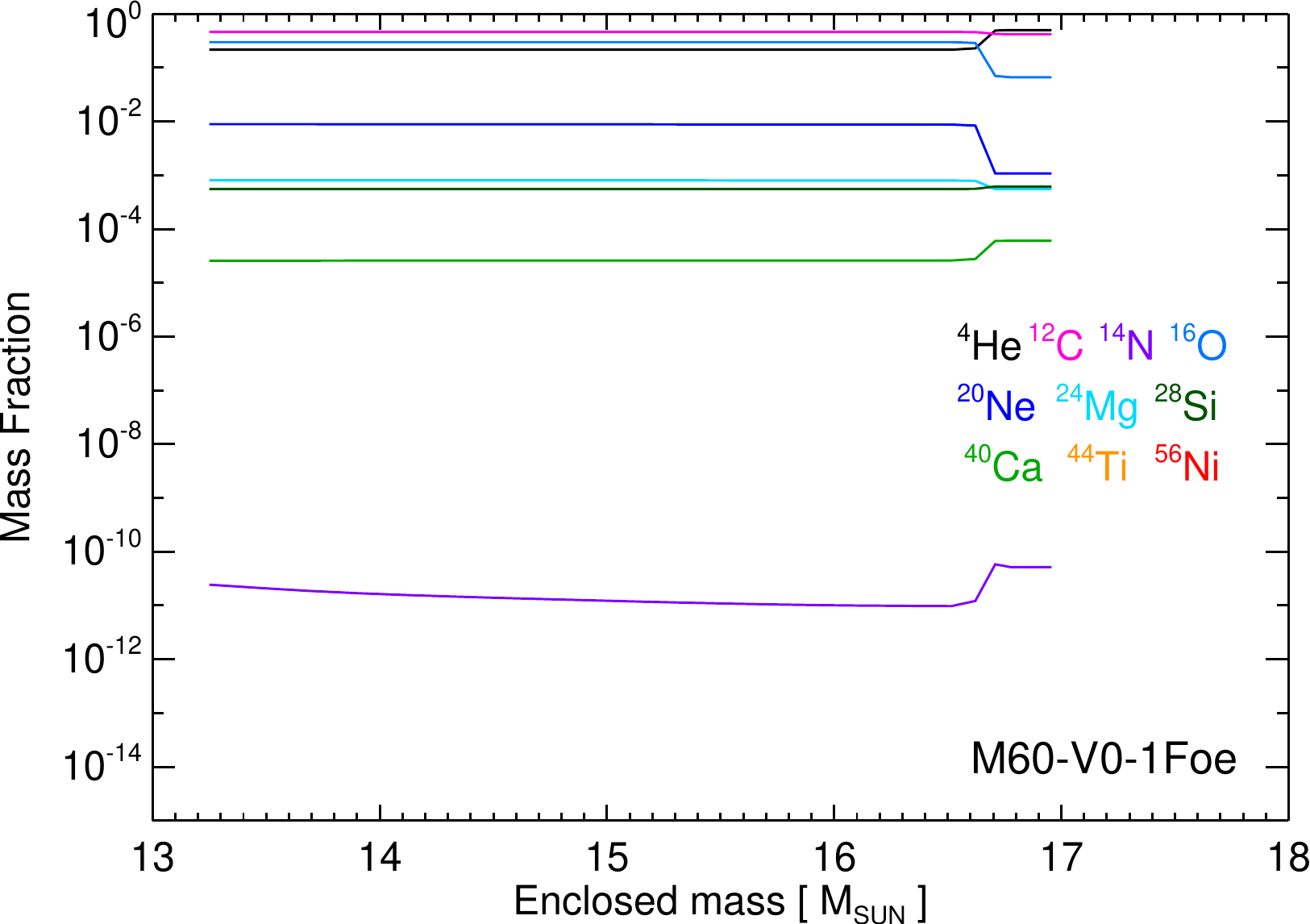}
    \includegraphics*[width=0.5\hsize]{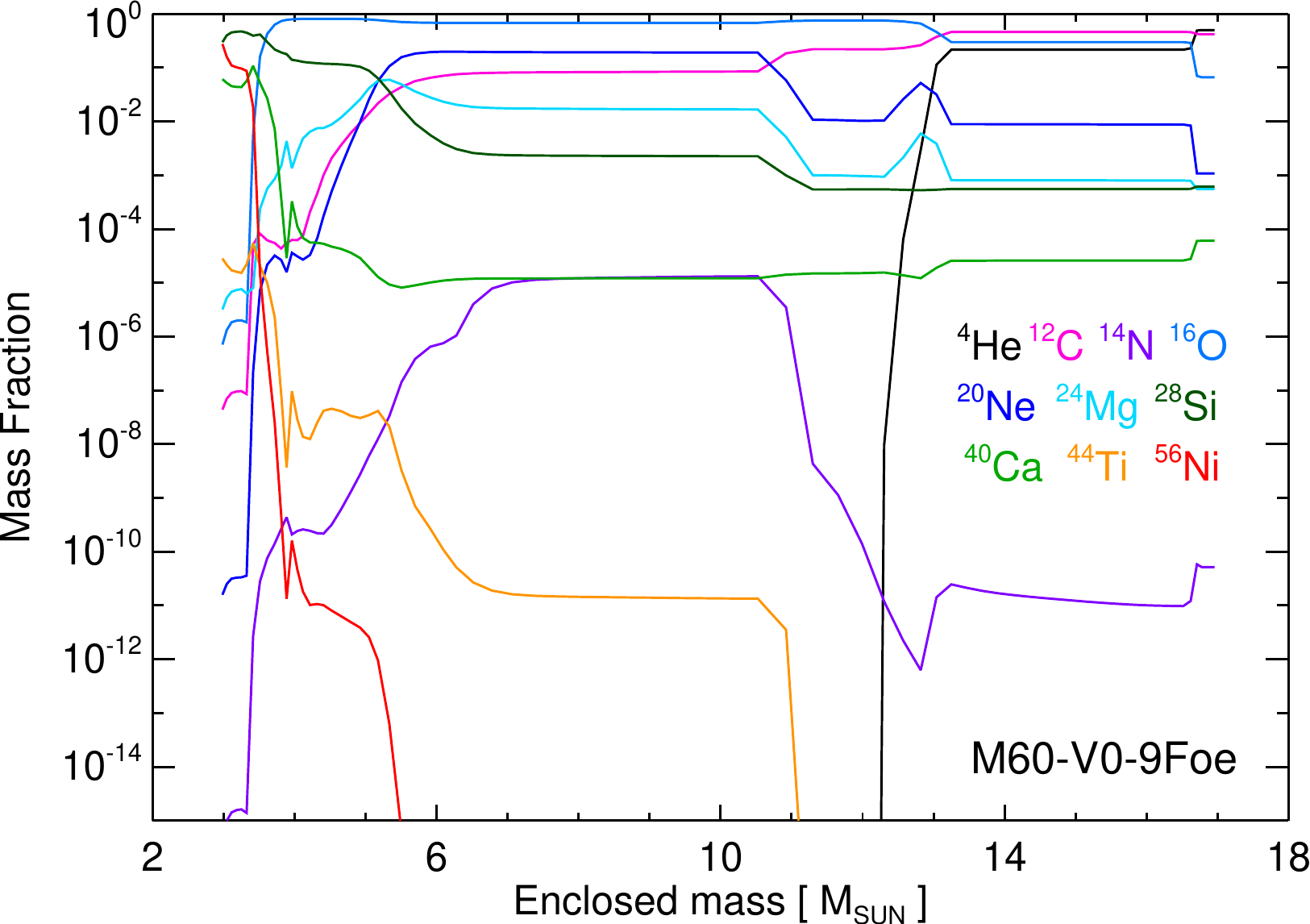}
    \caption{Left upper panel: pre-SN evolution of stars of 60~$M_\odot$ 
    and 80~$M_\odot$ (from the main sequence phase up to the core He 
    depletion stage, see \citealt{lim18}), with solar (initial) 
    metallicities and initial rotation velocities 
    $V_{\mathrm{rot}}=0$~km~s$^{-1}$ (solid lines) and 
    $V_{\mathrm{rot}}=300$~km~s$^{-1}$ (dotted lines).
    The yellow asterisk marks the position of G26 in the HR diagram.
    Right upper panel: initial radial profiles of density in the models 
    defined in Table~\ref{Tab:models}. Models with initial rotation velocities 
    $V_{\mathrm{rot}}=0$~km~s$^{-1}$ and $V_{\mathrm{rot}}=300$~km~s$^{-1}$ 
    are represented with solid and dotted lines respectively.
    Lower panels: initial radial profiles of mass fractions for the species considered 
    for two representative models, one with a low explosion energy (left panel) and the other with a high explosion energy (right panel). In both cases, the model describes a zero-age main-sequence star of 60 $M_{\odot}$.
    The range of enclosed mass ($x$-axis) is very different in these models
    due to the different location of the mass cut \citep[see][]{lim18}.}
    \label{Fig:progenitor}
  \end{figure*}

  We modeled the post-explosion evolution of a core-collapse SN starting 
  immediately after the shock breakout, and we followed the transition 
  from the SN to the SNR phase and the interaction of the remnant with 
  the inhomogeneous pre-SN environment. 
  As initial conditions, we adopted the explosive nucleosynthesis models described in \cite{lim18}. 
  These authors presented a grid of pre-SN models of massive stars 
  whose mass spans the range between $13$ and $120\,M_\odot$, covering four 
  metallicities (i.e., [Fe/H]=0, -1, -2, and -3) and three initial 
  rotation velocities (i.e., 0, 150, and 300~km~s$^{-1}$).
  In particular, we selected the models with solar metallicity and
  star masses either $60$ or $80\,M_\odot$, which reproduce evolutionary tracks in agreement with the position of G26 in the Hertzsprung-Russell (HR) diagram (see left upper panel in Fig.~\ref{Fig:progenitor}; see also 
  Fig.~14 in \citealt{lim18}, and Fig.~3 in \citealt{smi17}). 
  For the initial rotation velocity, we explored the models with the 
  two extreme values\footnote{For the 
  sake of completeness, we selected the stars with the most extreme values 
  of initial rotation velocity \citep[see][]{lim18}, even if the evolution of the models with $V_{\mathrm{rot}}=300$~km~s$^{-1}$ is not in good agreement with the position of G26 in the HR diagram (see Fig.~\ref{Fig:progenitor}).}, namely either 0 or 300~km~s$^{-1}$. 
  For completeness, for each case, we 
  selected a SN with either low or high explosion energy, considering the total kinetic
  energy of the ejecta\footnote{At this stage the energy 
  of the ejecta is almost entirely kinetic, being the internal energy 
  only a small percentage of the total energy.}.
  A summary of the cases explored is given in Table \ref{Tab:models}, where we report\footnote{The parameters outlined in Table~\ref{Tab:models} are referred
  to the values used for the pre-SN models described in \cite{lim18}.}: the main-sequence mass of the star, $M_*$; the initial rotation velocity of the star, $V_{\mathrm{rot}}$; the energy of the explosion, $E_{\rm exp}$.
  
  \begin{table}
    \caption[]{Summary of the pre-SN models \citep{lim18} 
    adopted as initial conditions.}
    \label{Tab:models}
    \centering
    \begin{tabular}{cccc}
    \hline\hline
    Model  &  $M_*$ ($M_\odot$)  &  $V_{\mathrm{rot}}$ (km~s$^{-1}$)  
    &  $E_{\rm exp}$ (erg) \\
    \hline
    M60-V0-1Foe  &  60  &  0  &  $1\times 10^{51}$ \\
    M60-V300-1Foe  &  60  &  300  &  $1\times 10^{51}$ \\
    M60-V0-9Foe  &  60  &  0  &  $9\times 10^{51}$ \\
    M60-V300-9Foe  &  60  &  300  &  $9\times 10^{51}$ \\
    \hline
    M80-V0-1Foe  &  80  &  0  &  $1\times 10^{51}$ \\ 
    M80-V300-1Foe  &  80  &  300  &  $1\times 10^{51}$ \\
    M80-V0-12Foe  &  80  &  0  &  $12\times 10^{51}$ \\
    M80-V300-12Foe  &  80  &  300  &  $12\times 10^{51}$ \\    
    \hline
    \end{tabular}
  \end{table}
  
  The initial blast wave is defined from the 1D profiles of density, 
  pressure, velocity, and abundances of 10 species\footnote{We do not 
  include the H in our model because its 
  total mass is very low in all the progenitors considered here (see 
  Table~\ref{Tab:models}), as it has been expelled by the star almost 
  totally into the CSM before the SN event.} ($^4$He, $^{12}$C, 
  $^{14}$N, $^{16}$O, $^{20}$Ne, $^{24}$Mg, $^{28}$Si, $^{40}$Ca, $^{44}$Ti, 
  $^{56}$Ni) describing the ejecta
  after the shock breakout. In Figure~\ref{Fig:progenitor}, we present 
  the profiles of density for the models outlined in Table~\ref{Tab:models}
  (top right panel), and the abundances of the 10 species considered for
  two reference cases, one with low (left bottom panel) and the other with high (right bottom 
  panel) explosion energy, and both for a zero-age main-sequence star of 60 $M_{\odot}$. The range of enclosed mass, plotted on the $x$-axis, is very different
  for low and high explosion energy models due to the different location of the mass cut \citep[see][]{lim18}.
  
  In our models, these 1D profiles are mapped in 
  the 3D domain, assuming spherical symmetry, and centered at the origin of the 3D Cartesian coordinate 
  system. We assumed a clumpy initial density structure of the ejecta, 
  as suggested by theoretical and spectropolarimetric studies (e.g., \citealt{nag00,kif06,wan03,wan04,wan08,gaw10,hol10,won15}). Thus, 
  after the 1D profiles of ejecta are remapped into the 3D domain, the 
  small-scale structure of the ejecta is modeled as per-cell random density distributions 
  by adopting a power-law probability distribution \citep[see][]{orl12}.
  In our simulations, the ejecta clumps have the same initial size (about 
  2\% of the initial remnant radius), and a maximum density perturbation
  $\nu_{\rm max}=10$. In Table~\ref{Tab:mass}, we outline the 
  total masses of the chemical elements composing the ejecta for all 
  the models presented in Table~\ref{Tab:models}. In Table~\ref{Tab:domain}, we report: the total mass of the fallback, $M_\mathrm{rem}$; the mass of the ejecta, $M_\mathrm{ej}$; the initial time of the simulation, $t_\mathrm{0}$; the radius of the sphere containing the ejecta at $t_\mathrm{0}$, $R_{\mathrm{ej}}$; the extension of the domain, in the first, $D_{\mathrm{i}}$, and last, $D_{\mathrm{f}}$, remapping (see later).
  
  \begin{table*}
    \caption[]{Summary of the total masses (in units of $M_\odot$) of the chemical elements composing the ejecta in the models described in Table~\ref{Tab:models}.}
    \label{Tab:mass}
    \centering
    \begin{tabular}{ccccccccccc}
    \hline\hline
    Model  &  $M_\mathrm{^4He}$  &  
    $M_\mathrm{^{12}C}$  &  $M_\mathrm{^{14}N}$  &  $M_\mathrm{^{16}O}$  &  
    $M_\mathrm{^{20}Ne}$  &  $M_\mathrm{^{24}Mg}$  &  $M_\mathrm{^{28}Si}$  &  
    $M_\mathrm{^{40}Ca}$  &  $M_\mathrm{^{44}Ti}$  &  $M_\mathrm{^{56}Ni}$ \\
    \hline
    M60-V0-1Foe & $0.78$ & 1.49 & $5.2\cdot10^{-11}$ & $0.92$ & 
    $0.027$ & $2.6\cdot10^{-3}$ & $1.8\cdot10^{-3}$ & 
    $9.3\cdot10^{-5}$ & $6.4\cdot10^{-34}$ & $6.9\cdot10^{-44}$ \\
    M60-V300-1Foe & $0.52$ & 1.15 & $1.0\cdot10^{-5}$ & 1.07 & 
    $0.087$ & $0.010$ & $1.6\cdot10^{-3}$ & 
    $6.4\cdot10^{-5}$ & $1.2\cdot10^{-33}$ & $4.0\cdot10^{-43}$ \\ 
    M60-V0-9Foe & $0.89$ & 2.68 & $5.3\cdot10^{-5}$ & 7.84 & 
    1.16 & $0.15$ & $0.50$ & 
    $0.038$ & $1.6\cdot10^{-5}$ & $0.042$ \\
    M60-V300-9Foe & $0.62$ & 2.20 & $4.5\cdot10^{-5}$ & 8.04 & 
    1.30 & $0.17$ & $0.45$ & 
    $0.031$ & $1.2\cdot10^{-5}$ & $0.026$ \\
    \hline
    M80-V0-1Foe & $0.69$ & 1.14 & $2.4\cdot10^{-10}$ & $0.75$ & 
    $0.084$ & $0.012$ & $1.5\cdot10^{-3}$ & 
    $6.8\cdot10^{-5}$ & $1.0\cdot10^{-34}$ & $1.4\cdot10^{-43}$ \\ 
    M80-V300-1Foe & $0.59$ & 1.25 & $4.0\cdot10^{-6}$ & 1.05 & 
    $0.072$ & $7.6\cdot10^{-3}$ & $1.6\cdot10^{-3}$ & 
    $6.9\cdot10^{-5}$ & $1.7\cdot10^{-33}$ & $5.2\cdot10^{-43}$ \\
    M80-V0-12Foe & 1.06 & 3.22 & $2.5\cdot10^{-5}$ & 11.4 & 
    1.20 & $0.19$ & $0.67$ & 
    $0.046$ & $1.9\cdot10^{-5}$ & $0.024$ \\ 
    M80-V300-12Foe & $0.70$ & 2.46 & $4.5\cdot10^{-5}$ & 8.43 & 
    1.15 & $0.16$ & $0.59$ & 
    $0.054$ & $2.7\cdot10^{-4}$ & $0.72$ \\ 
    \hline
    \end{tabular}
  \end{table*}
  
  \begin{table*}
    \caption[]{Summary of the parameters describing the initial 
    condition and the computational domain adopted for the models 
    described in Table~\ref{Tab:models}.}
    \label{Tab:domain}
    \centering
    \begin{tabular}{ccccccc}
    \hline\hline
    Model  &  $M_\mathrm{rem}$ ($M_\odot$)  &  $M_\mathrm{ej}$ ($M_\odot$)  &  
    $t_\mathrm{0}$ (s)  &  $R_{\mathrm{ej}}$ (cm)  &  
    $D_{\mathrm{i}}$ (cm)  &  $D_{\mathrm{f}}$ (cm)\\
    \hline
    M60-V0-1Foe  &  13.69  &  3.73  &  540  &  $\approx 9.5\times 10^{11}$  &  
    $\approx 2.3\times 10^{12}$  &  $\approx 1.6\times 10^{20}$ \\
    M60-V300-1Foe  &  13.75  &  2.91  &  37510  &  $\approx 1.1\times 10^{14}$  &  
    $\approx 2.6\times 10^{14}$  &  $\approx 1.6\times 10^{20}$ \\ 
    M60-V0-9Foe  &  3.04  &  13.41  &  500  &  $\approx 8.7\times 10^{12}$  &  
    $\approx 2.1\times 10^{13}$  &  $\approx 2.8\times 10^{20}$ \\
    M60-V300-9Foe  &  3.25  &  12.88  &  30794  &  $\approx 1.6\times 10^{14}$  &  
    $\approx 3.9\times 10^{14}$  &  $\approx 2.8\times 10^{20}$ \\
    \hline
    M80-V0-1Foe  &  20  &  3.33  &  610  &  $\approx 1.1\times 10^{12}$  &  
    $\approx 2.6\times 10^{12}$  &  $\approx 1.8\times 10^{20}$ \\ 
    M80-V300-1Foe  &  14.46  &  3  &  27818  &  $\approx 7.9\times 10^{13}$  &  
    $\approx 1.9\times 10^{14}$  &  $\approx 1.7\times 10^{20}$ \\
    M80-V0-12Foe  &  4.14  &  18.26  &  550  &  $\approx 1.9\times 10^{12}$  &  
    $\approx 4.5\times 10^{12}$  &  $\approx 2.6\times 10^{20}$ \\
    M80-V300-12Foe  &  2.39  &  14.58  &  33885  &  $\approx 2.0\times 10^{14}$  &  
    $\approx 4.7\times 10^{14}$  &  $\approx 2.8\times 10^{20}$ \\    
    \hline
    \end{tabular}
  \end{table*}
   
\begin{figure}
    \includegraphics*[width=\hsize]{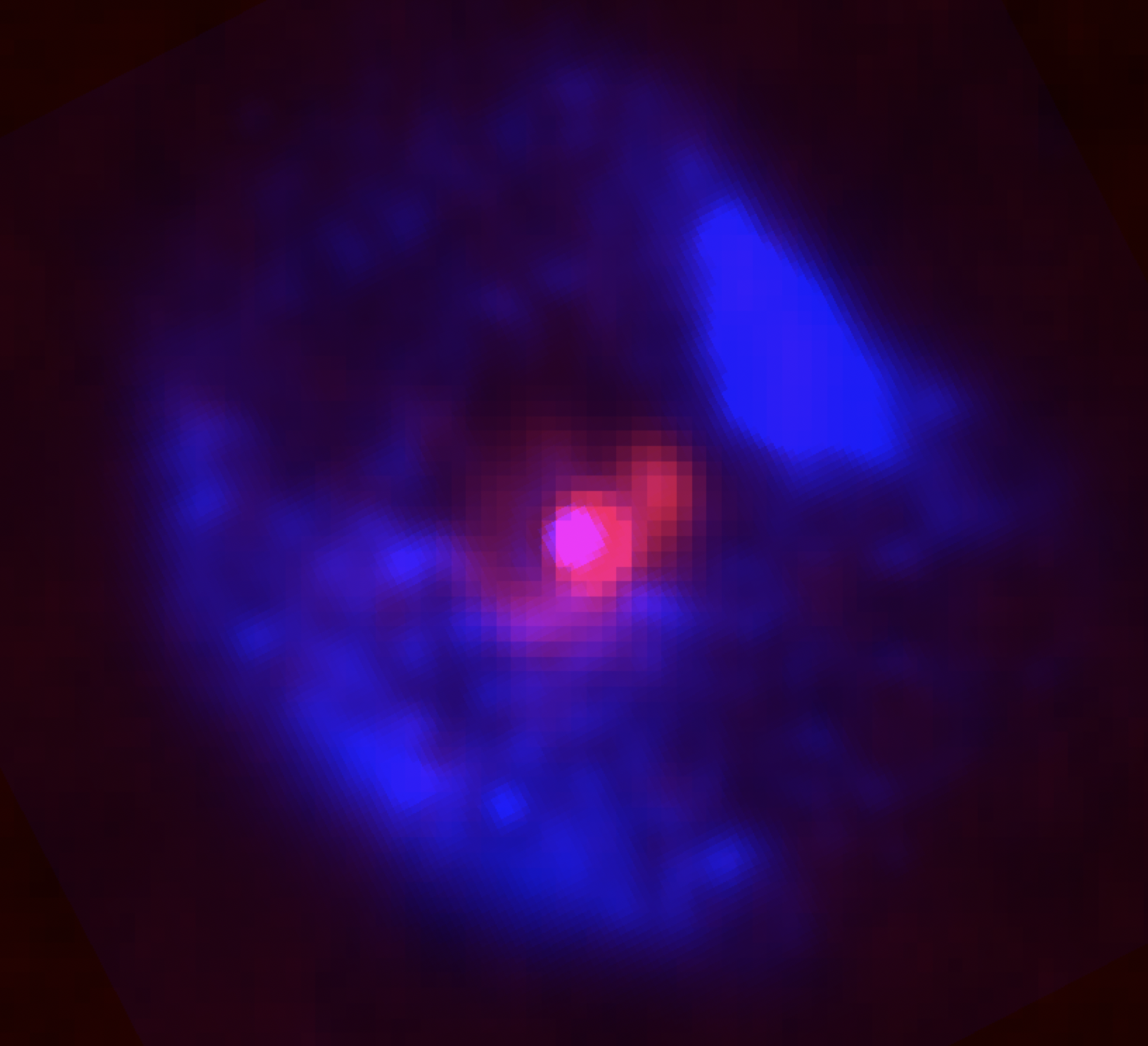}\\
    \includegraphics*[width=\hsize]{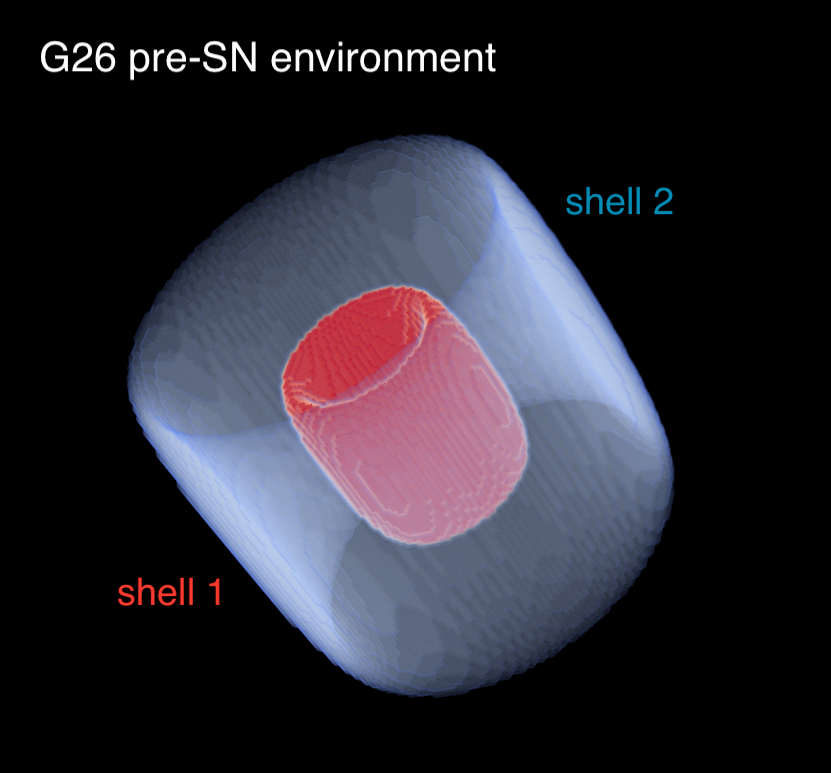}
    \caption{Top: Map of G26 which combines infrared at $24\, \mu m$ (blue) and 
    radio (red) images \citep[see][]{uma12}. 
    Bottom: Schematic view of the CSM around G26, 
    representing the two dense shells (in red and light blue) observed in the radio and infrared bands. The LBV is located in the center of the two nested 
    toroidal shells.}
    \label{Fig:initial}
  \end{figure}

  The pre-SN environment is parametrized by following the two shells model
  proposed for G26 by \cite{uma12} based on radio and infrared observations
  (see Fig.~\ref{Fig:initial}; see also Fig.~2 in \citealt{uma12}). 
  According to \cite{uma12}, the very massive nebula of G26 consists
  of, at least, $17\,M_\odot$ of ionized gas\footnote{This has to be 
  considered as a lower limit of the true content in mass of the entire nebula 
  \citep[see][]{uma12}}, divided into $7.6\,M_\odot$ and $9.7\,M_\odot$ 
  in the inner and outermost nebula respectively. The estimated total mass of dust is $1.2-3.2\times 10^{-2}\,M_\odot$, which is several orders of magnitude lower than the mass of the ionized gas and thus negligible for the purposes of this paper. In the light of this,
  we defined an ambient medium consisting of a spherically symmetric steady wind 
  and two dense nested shells (representing the inner and outermost nebula 
  observed) with a common axis coincident with the $z$-axis (see bottom panel in 
  Fig.~\ref{Fig:initial}). We assumed the two shells (possibly related to different 
  mass loss episodes occurred in the past) each with a mass of $\sim 10 \,M_\odot$, namely slightly higher than the two lower limits ($7.6\,M_\odot$ and $9.7\,M_\odot$) found by \citealt{uma12}).
  The spherically symmetric wind is characterized by a gas density 
  proportional to r$^{-2}$, defined following a mass loss rate\footnote{The mass-loss rate has 
  been derived assuming a wind velocity of 200~km~s$^{-1}$, which is 
  within the range observed for LBVs and LBV candidates 
  (i.e. $100-250$~km~s$^{-1}$; \citealt{vin12}).} 
  $\dot M$=10$^{-4}\,M_\odot$~yr$^{-1}$. 
  We fixed a lower threshold of 0.1~cm$^{-3}$ for the pre-SN density of the CSM, which corresponds to assume a progressive flattening of the wind profile to a uniform density at large radii\footnote{The flattening of the density profile is introduced to prevent unrealistic low values of the wind density that would be otherwise described with the $r^{-2}$ profile.}.
  We note that the environment outlined in the bottom panel in Fig.~\ref{Fig:initial}
  is a simplified version of the CSM expected in a LBV exploding star. 
  In particular, we expect to find a more complex and dense CSM close to the star.
  However, due to the lack of information about the more internal CSM, here we focus on investigating the effect that a SN interacting with the toroidal structures identified in G26 by \cite{uma12} could have.
  
  The two shells are defined as clumpy structures azimuthally symmetric 
  about the $z$-axis, centered at the origin of the 3D Cartesian coordinate 
  system $(x_\mathrm{0}, y_\mathrm{0}, z_\mathrm{0}) = (0,0,0)$. 
  They follow the shape of an elliptic torus, a surface of revolution 
  produced by rotating an ellipse, in this case,
  about the $z$-axis. The dimensions of the shells were chosen following 
  the results of \cite{uma12}. The radius of the inner and outer shells 
  (i.e., the distances from the center of explosion to the center of the ellipses) 
  are $\approx0.6$~pc and $\approx1.6$~pc respectively. Assuming a wind velocity of 200 km s$^{-1}$ (e.g.,~\citealt{smi17}), the shells would be the relics of mass ejection episodes occurred between 3000 and 8000 years before the SN event. The length 
  of the major semi-axis (along the $z$-direction) is 0.8 and 1.5~pc for the inner and 
  outer shells, respectively; the size of the minor semi-axis (in the $xy$-plane) is 
  0.15~pc in both cases.
  
  The two dense shells are expected to play a central role in modifying the expansion 
  of the forward shock and in driving a reflected shock through the ejecta. 
  Since the geometry and density distribution adopted in the paper are 
  idealized, this may introduce some features in the remnant structure 
  if the shells are assumed to be uniform. In order to get a non-uniform density distribution for the two shells, 
  the material was modeled as a set of spherical clumps with radius 
  $\approx$~0.15~pc randomly distributed, filled with spherical sub-clumps of radius $\approx$~0.05~pc (see, for instance, \citealt{ust21}).
  The density of the plasma in the spheres follows a normal distribution
  with mean density 160~cm$^{-3}$ and 45~cm$^{-3}$ for the inner and 
  outer shells respectively, in agreement with the values of average 
  electron density estimated by \cite{uma12} for the corresponding
  shells in G26 (assuming a distance of 6.5~kpc, see \citealt{cla03}). 
  The total mass in each modeled shell is $\approx 12 M_\odot$, in agreement with the estimates given by \cite{uma12}. 
  We assume the two dense shells to be in pressure equilibrium with the environment. 
  Note that we neglected any bulk velocity in both components of the pre-SN CSM, namely the 
  wind component and the two dense shells, because it would be anyway much smaller 
  than the velocity of the forward shock. Furthermore, due to lack of observational constraints and evidence, we neglected mass eruptions that may have occurred in the latest phases of the progenitor evolution before core-collapse and that could have generated a dense and inhomogeneous medium in the immediate surrounding of the SN. In fact, we have no indications of this material from observations of G26, and we preferred to keep the CSM as simple as possible to investigate the effects of the extended shells observed in G26 on the SNR evolution.
  % Discorso masse totali - wind bubble?
   
  The simulations include passive tracers ($C$) to follow the evolution of the
  different plasma components (the ejecta and the two dense shells), 
  to store information on the shocked plasma (time, shock velocity, 
  and shock position when a cell of the 
  mesh is shocked by either the forward or the reverse shock), and to 
  follow the chemical evolution of the ejecta for 10 different species 
  ($^4$He, $^{12}$C, $^{14}$N, $^{16}$O, $^{20}$Ne, $^{24}$Mg, $^{28}$Si, 
  $^{40}$Ca, $^{44}$Ti, $^{56}$Ni).
  The continuity equations of the tracers are solved in addition to 
  our set of HD equations. In the case of tracers associated with 
  the different plasma components (ejecta or the shells), each material is initialized with 
  $C_{\mathrm{i}} = 1$, while $C_{\mathrm{i}} = 0$ elsewhere, where the 
  index $i$ refers to the ejecta as a whole, or to the material in the inner or in outer shell. 
  The chemical evolution of the ejecta is followed by adopting a multiple 
  fluids approach \citep[see e.g.][]{orl16,orl21}. The fluids following 
  the evolution of the different species are initialized with the 
  abundances calculated at the shock breakout with the SN models 
  described in \cite{lim18} and calculated from the set of stellar models summarized in Table~\ref{Tab:models} (see Figure~\ref{Fig:progenitor}).
  The different fluids mix together during the evolution and, in particular, 
  when the ejecta interact with the reverse shock that develops during the 
  expansion of the remnant. The density of a specific element in a fluid
  cell is calculated as $\rho_{\mathrm{i}}=\rho \cdot C_{\mathrm{i}}$, 
  where, in this case, $C_{\mathrm{i}}$ is the mass fraction of each 
  element and the index $i$ refers to the considered element.
  This approach allows us to follow the spatial distribution of 
  the chemical elements both inside and outside the reverse shock 
  during the model evolution. All the other tracers (in particular those which store information on the shocked plasma) are initialized to zero everywhere.
  
  The computational domain is a Cartesian box covered by a uniform grid
  of $512\times512\times512$ zones, including the initial remnant defined by the profiles presented
  in Fig.~\ref{Fig:progenitor} \citep[see][]{lim18} mapped in 3D.
  The initial computational domain extends from $(-D_{\mathrm{i}}/2)$ to
  $(D_{\mathrm{i}}/2)$ in all the directions (see Table~\ref{Tab:domain}), 
  to cover the sphere containing the ejecta at the beginning of the 
  simulation, leading to a spatial resolution in the range 
  $4.5\times 10^9-9.2\times 10^{11}$~cm depending on the model. 
  In order to follow the large physical scales spanned during the remnant 
  expansion, we adopted the approach described in \cite{orl19,orl20}.
  During the evolution, the computational domain has been gradually extended 
  following the expansion of the remnant through the CSM and the physical 
  quantities have been remapped in the new domain. The domain is extended 
  by a factor of 1.2 in all directions when the forward shock reaches 
  one of the boundaries of the Cartesian box. The number of mesh points 
  is the same at each remapping, thus the spatial resolution gradually 
  decreases during the evolution. In each remapping, all the physical quantities in the
  new region, added outside the previous computational domain, are set to the values of the pre-SN CSM. For the models explored here, a number between 74 and 100 remappings 
  have been necessary to follow the interaction of the remnant with the CSM during 10000~yr of evolution. The final domain extends between $(-D_{\mathrm{f}}/2)$ and 
  $(D_{\mathrm{f}}/2)$ in all the directions (see Table~\ref{Tab:domain}),
  leading to a spatial resolution in the range 
  $3.0\times 10^{17}-5.5\times 10^{17}$~cm depending on the model.
  All physical quantities were fixed to the values of the
  pre-SN CSM at all boundaries.

%--------------------------------------------------------------------
\section{Results}
\label{Sec:results}

\subsection{Hydrodynamic evolution}
\label{Sec:hd}

  Soon after the shock breakout, the ejecta propagate freely through the spherically symmetric wind, driving a forward shock in the wind and a reverse shock backward through the ejecta. During this phase of evolution, the unshocked ejecta expand almost homologously, thus maintaining their initial 
  structure and chemical stratification. After $\approx 18-40$~yr of evolution, depending on the
  case, the forward shock hits the innermost dense shell of the CSM (see Fig.~\ref{Fig:initial}). In this section, we describe
  the interaction of the modeled remnants (see Table~\ref{Tab:models}) with the two shells and their subsequent evolution for $\approx 10000$
  years.
  In all the figures presented in this section, we assumed the system oriented as the toroidal shells of G26, as deduced from the analysis of observations \citep[see Fig. 2 in ][]{uma12}.
  To this end, we rotated the original system about the three axes by the angles $i_x=30$º, $i_y=30$º and $i_z=25$º, to fit the orientation of the shells in G26 with respect to the line of sight (LoS).
  
  \begin{figure*}
    \includegraphics*[width=\hsize]{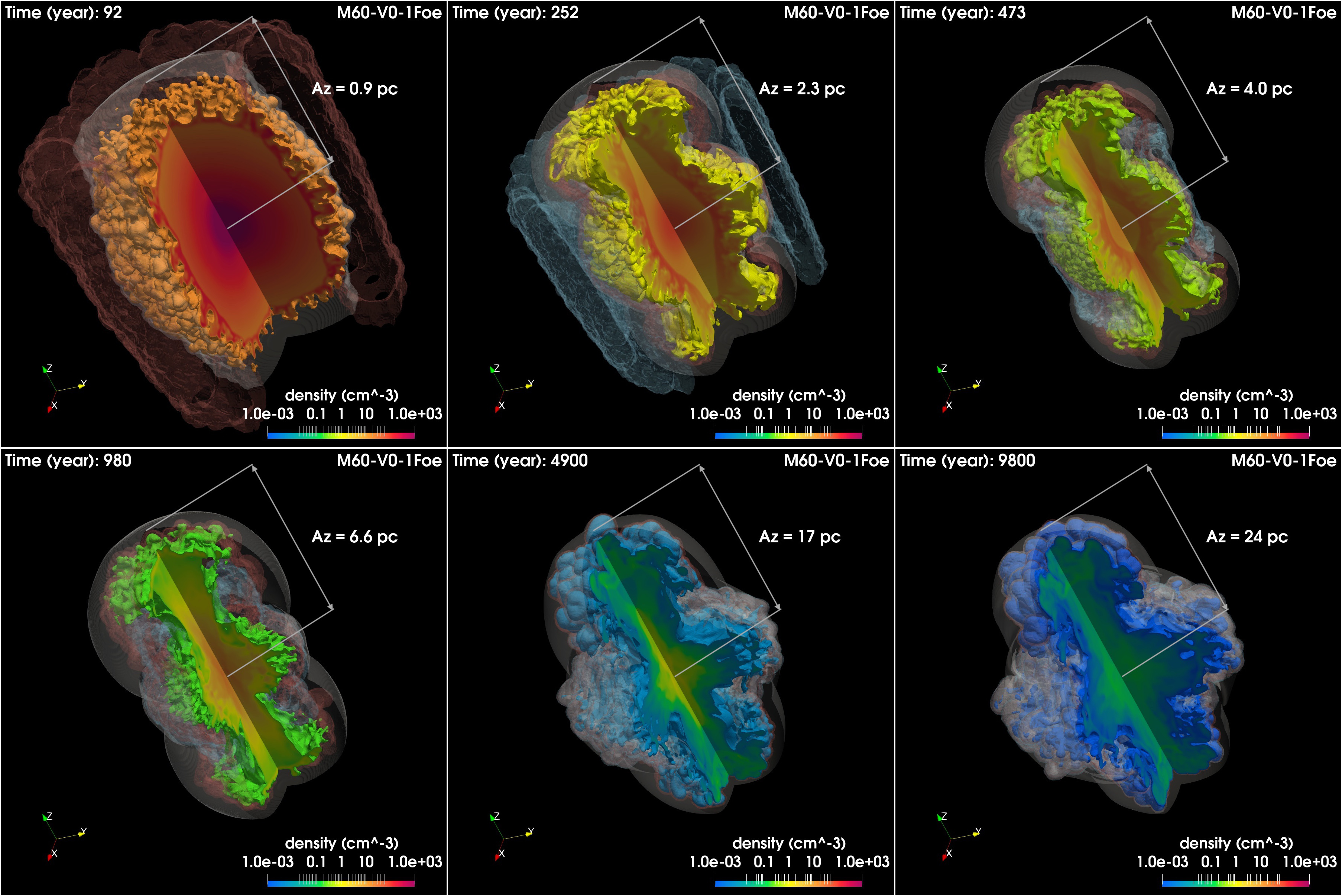}
    \caption{Density distributions of the ejecta of the model M60-V0-1Foe 
    at different evolution times (increasing from upper left to lower right).
    The opaque irregular isosurfaces correspond to a value of density
    at 1\% of the peak density with one quadrant cut in order to see 
    the radial distribution. The semi-transparent surface marks the 
    position of the forward shock; the initially toroidal semi-transparent 
    structures in red and cyan colors represent the inner and outer shells 
    in the CSM respectively. The system is oriented as G26 (see 
    Fig.~\ref{Fig:initial}), corresponding to the rotation angles $i_x=30$º, 
    $i_y=30$º, $i_z=25$º about the $x$, $y$, and $z$ axes, respectively. 
    The evolution time is shown in the upper left corner of each panel. 
    $A_{\rm z}$ indicates the distance of the forward shock from the 
    center of the explosion along the z-axis.
    The complete temporal evolution is available as online 
    movie (Movie~1).}
    \label{Fig:hydro}
  \end{figure*}
  
\begin{figure*}
    \includegraphics*[width=\hsize]{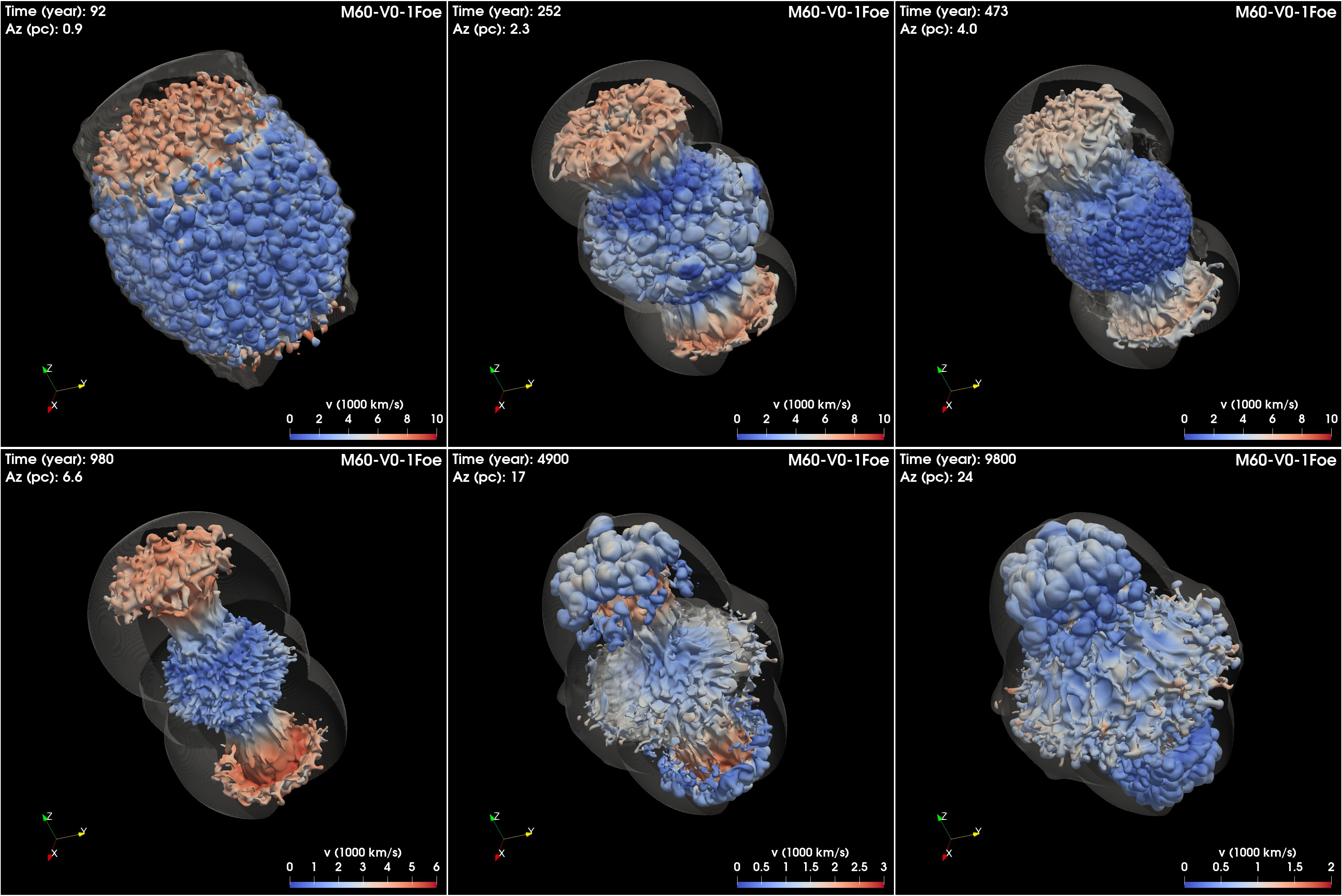}
    \caption{Same as in Fig.~\ref{Fig:hydro} but the colors give the radial 
    velocity in units of 1000~km~s$^{-1}$ on the isosurface.
    The complete temporal evolution is available as online movie (Movie~2).}
    \label{Fig:vel}
  \end{figure*}

  The evolution of model M60-V0-1Foe is shown in Figs.~\ref{Fig:hydro}
  and \ref{Fig:vel}, at different epochs (increasing from upper left to 
  lower right panels). Fig.~\ref{Fig:hydro} reports the density 
  distributions for the ejecta in logarithmic scale; Fig.~\ref{Fig:vel}
  shows the same isosurfaces that appear in Fig.~\ref{Fig:hydro}, but colored 
  according to the corresponding radial velocity (note the different 
  color scale in the lower panels). 
  The complete temporal evolution is available as online movies (Movie~1 
  and Movie~2). The total mass of the 
  ejecta in this model is low, namely $\approx 4\, M_\odot$ (see Table~\ref{Tab:domain}), as most of the stellar mass was lost in the 
  CSM during the pre-SN phases of the star, and in the fallback during 
  the core-collapse event ($\approx 14\, M_\odot$, see Table~\ref{Tab:domain}). The forward shock, whose position is indicated with a 
  semi-transparent surface in both figures, starts to interact with the 
  innermost dense shell (red clumpy semi-transparent structure in the top 
  left panel in Fig.~\ref{Fig:hydro}) after $\sim 40$~yr of evolution. 
  The interaction determines a strong slowdown of the forward shock which propagates through the shell (see top left panel in Fig.~\ref{Fig:vel}), 
  and, consequently, a strengthening of the reverse shock traveling 
  through the ejecta. Meanwhile, the forward shock continues to expand 
  freely through the wind along the polar directions (roughly along the $z$-axis). 
  As a result, the initial quasi-spherical distribution 
  of ejecta progressively becomes asymmetric (elongated in the $z$
  direction) in the subsequent evolution (see top central panel in 
  Figs.~\ref{Fig:hydro}~and~\ref{Fig:vel}). In a few years, at t~$\approx 115$~yr, the forward shock goes beyond the inner shell in the equatorial plane and starts traveling again through the wind. After $\approx 250$ years of evolution, the blast hits the outer dense shell (cyan semi-transparent clumpy 
  structure in the top middle panel in Fig.~\ref{Fig:hydro}). 
  Similarly to the previously encountered shell, the dense toroidal 
  structure slows down the expansion of the forward shock (see top right panel in Fig.~\ref{Fig:vel}), 
  and, again, it produces a strengthening of the reverse shock traveling 
  through the ejecta. This together with the free expansion of the ejecta at the poles enhance even more the elongated shape of the remnant 
  (see top right panel in Fig.~\ref{Fig:hydro}). The reverse shock, 
  powered by the interaction with the two dense shells, heats the 
  internal ejecta and, after $\approx 1000$~yr of evolution, refocuses 
  approximately on the $z$-axis in proximity of the center of the explosion (see bottom left panel
  in Fig.~\ref{Fig:hydro}). The almost perfect refocusing of the reverse shock is enhanced by the idealized cylindrical symmetry of the system and, in particular of the two shells adopted. At this point, the central part of the 
  ejecta are confined by the material of the shells, while the terminal 
  edges along the poles continue to expand at high velocity 
  (see bottom left panel in Fig.~\ref{Fig:vel}).
  After the interaction with the two shells, the remnant continues 
  to expand through the wind of the progenitor star 
  (see bottom middle panels in Figs.~\ref{Fig:hydro}~and~\ref{Fig:vel}).
  At the end of the simulation, namely after $\approx 10000$~yr of 
  evolution, the ejecta have slowed down their expansion (see bottom 
  right panel in Fig.~\ref{Fig:vel}) and they show an elongated 
  shape due to the interaction with the two dense toroidal shells. At this time, the morphology is characterized by a broad jet-like structure with maximum density along the $z$-axis, 
  which extends for $\approx 24$~pc from the center of the explosion 
  (see bottom right panel in Fig.~\ref{Fig:hydro}). It is worth to note that, in our simulations, we assumed that the remnant propagates through an almost uniform ambient environment at large distances from the center of explosion (namely, due to the flattening of the wind density profile to 0.1~cm$^{-3}$; see Sect.~\ref{Sec:model}). However, at these distances, the remnant is expected to propagate through an inhomogeneous ISM that may partially wash out the fingerprints of the previous interaction of the remnant with the two dense shells.
  
  \begin{figure*}
    \includegraphics*[width=\hsize]{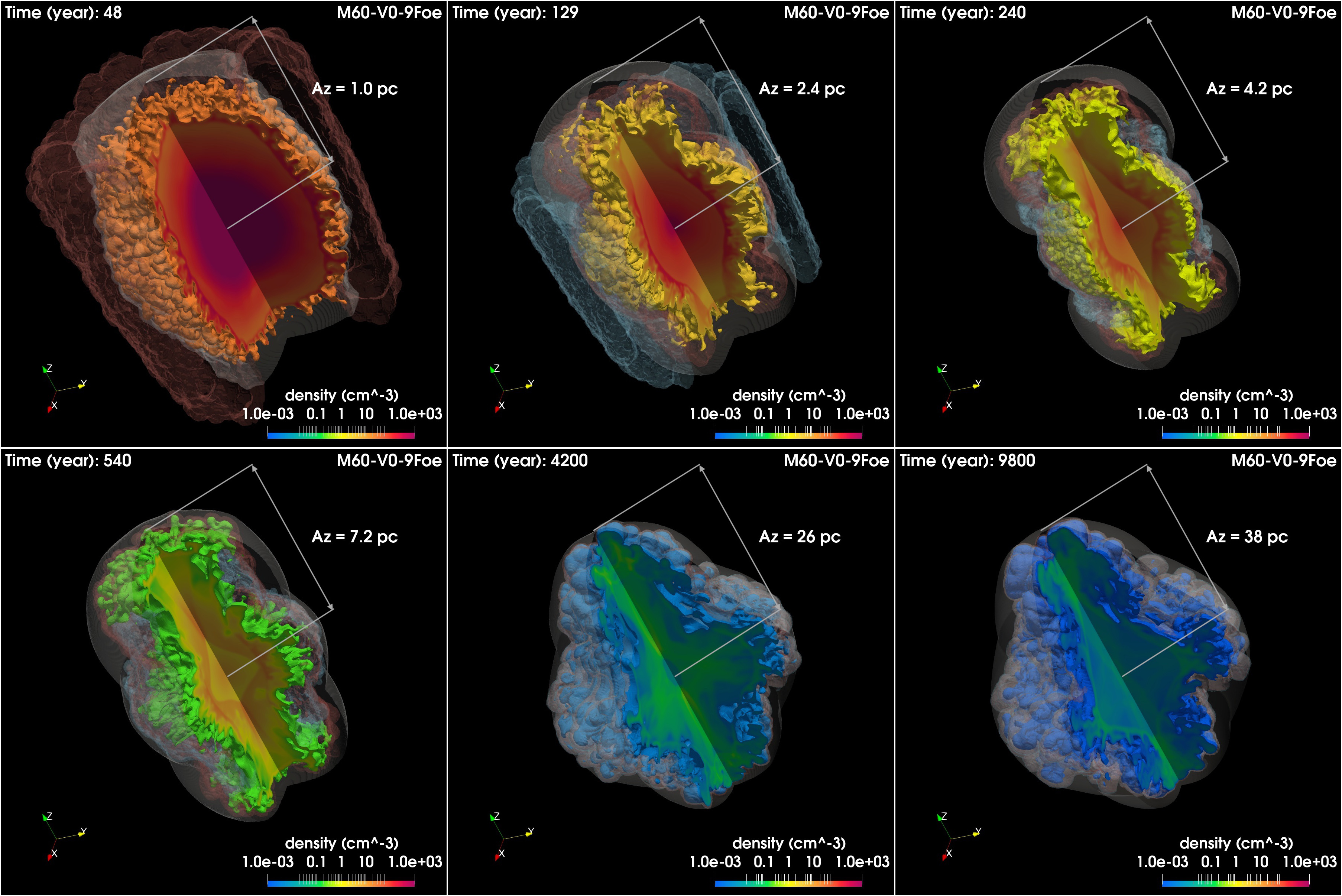}
    \caption{Same as in Fig.~\ref{Fig:hydro} but for model M60-V0-9Foe. The complete temporal evolution is available as online movie (Movie~3).}
    \label{Fig:hydro_en}
  \end{figure*}
  
  \begin{figure*}
    \includegraphics*[width=\hsize]{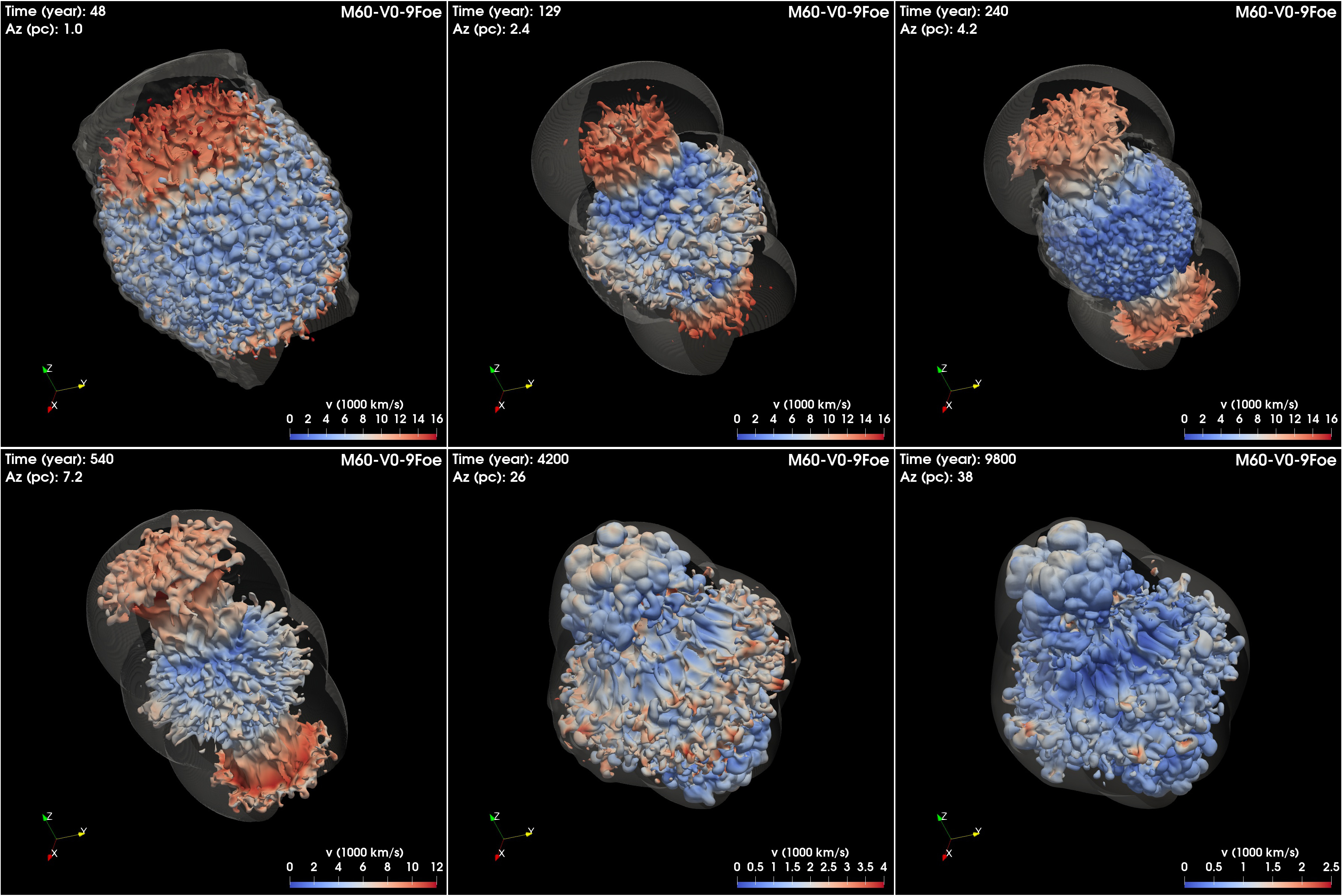}
    \caption{Same as in Fig.~\ref{Fig:hydro_en} but the colors give the radial 
    velocity in units of 1000~km~s$^{-1}$ on the isosurface.
    The complete temporal evolution is available as online movie (Movie~4).}
    \label{Fig:vel_en}
  \end{figure*}
  
  Similarly to the case with low explosion energy reported in Figs.~\ref{Fig:hydro}
  and ~\ref{Fig:vel}, the evolution of model M60-V0-9Foe is shown 
  in Figures~\ref{Fig:hydro_en}~and~\ref{Fig:vel_en}, at different 
  epochs (increasing from upper left to lower right panels).
  We note the different scales used in these figures in comparison
  with the analogous ones for model M60-V0-1Foe.
  The complete temporal evolution is available as online movies 
  (Movie~3 and Movie~4). The total mass of the 
  ejecta in model M60-V0-9Foe is much higher than in model M60-V0-1Foe, namely $\approx 13\, M_\odot$, as only $\approx 3\, M_\odot$ 
  were lost in the fallback during the core-collapse event (see 
  Table~\ref{Tab:domain}).
  In this case, the forward shock moves fastly through the CSM powered 
  by the high kinetic energy of the ejecta and starts to interact much earlier
  with the inner dense shell (red clumpy semi-transparent structure in the top left panels in Figs.~\ref{Fig:hydro} and~\ref{Fig:hydro_en}), 
  namely after $\sim 20$~yr of evolution. 
  As in model M60-V0-1Foe, the interaction determines a slowdown of the forward 
  shock which travels through the shell (see top left panel in 
  Fig.~\ref{Fig:vel_en}) and a strengthening of the reverse shock 
  traveling through the ejecta, but in this case the effect is far 
  less pronounced due to the high kinetic energy of the ejecta (see 
  top left and middle panels in Figs.~\ref{Fig:hydro},~\ref{Fig:vel},
  ~\ref{Fig:hydro_en}~and~\ref{Fig:vel_en}). The interaction of the remnant with the shell lasts for $\sim 38$ years. Then the forward shock, represented with a semi-transparent surface in 
  Fig.~\ref{Fig:hydro_en}, starts traveling again through the $r^{-2}$ wind of the progenitor. After $\approx 130$~yr of evolution, the blast wave hits the outer shell (cyan semi-transparent clumpy structure in the top middle 
  panel in Fig.~\ref{Fig:hydro_en}). Also in this case, the dense 
  shell squeezes the ejecta and pushes them along the $z$-direction (see
  top right panels in Figs.~\ref{Fig:hydro_en}~and~\ref{Fig:vel_en}). 
  However, due to the high kinetic energy of the ejecta, the slow down of the forward shock through the dense shell is not as relevant as in model M60-V0-1Foe and the blast goes quickly beyond the shell (see bottom panels in Fig.~\ref{Fig:vel_en}). As a consequence, the relative strengthening of the reverse shock is not as evident as in model M60-V0-1Foe (see top right panel in Fig.~\ref{Fig:hydro}), and the shock refocuses on the $z$-axis much later. 
  For instance, at $t=4200$~yr the remnant has already reached a maximum expansion of $\approx26$~pc from the center (see bottom 
  middle panel in Fig.~\ref{Fig:hydro_en}) in model M60-V0-9Foe;  in other words, the remnant is more extended than 
  that of the analogous model with low explosion energy (M60-V0-1Foe) at the 
  end of the simulation (see bottom right panel in Fig~\ref{Fig:hydro}). However, at the same age, the reverse shock in model M60-V0-9Foe has not refocused and the inner ejecta have 
  not been heated yet. At the end of the simulation, namely at 
  $\approx 10000$~yr, the ejecta have an elongated shape forming a broad
  jet-like structure with maximum density along the $z$-axis, which 
  extends for $\approx 38$~pc from the center of the explosion 
  (see bottom right panel in Fig~\ref{Fig:hydro_en}).
  
  \begin{figure*}
    \includegraphics*[width=\hsize]{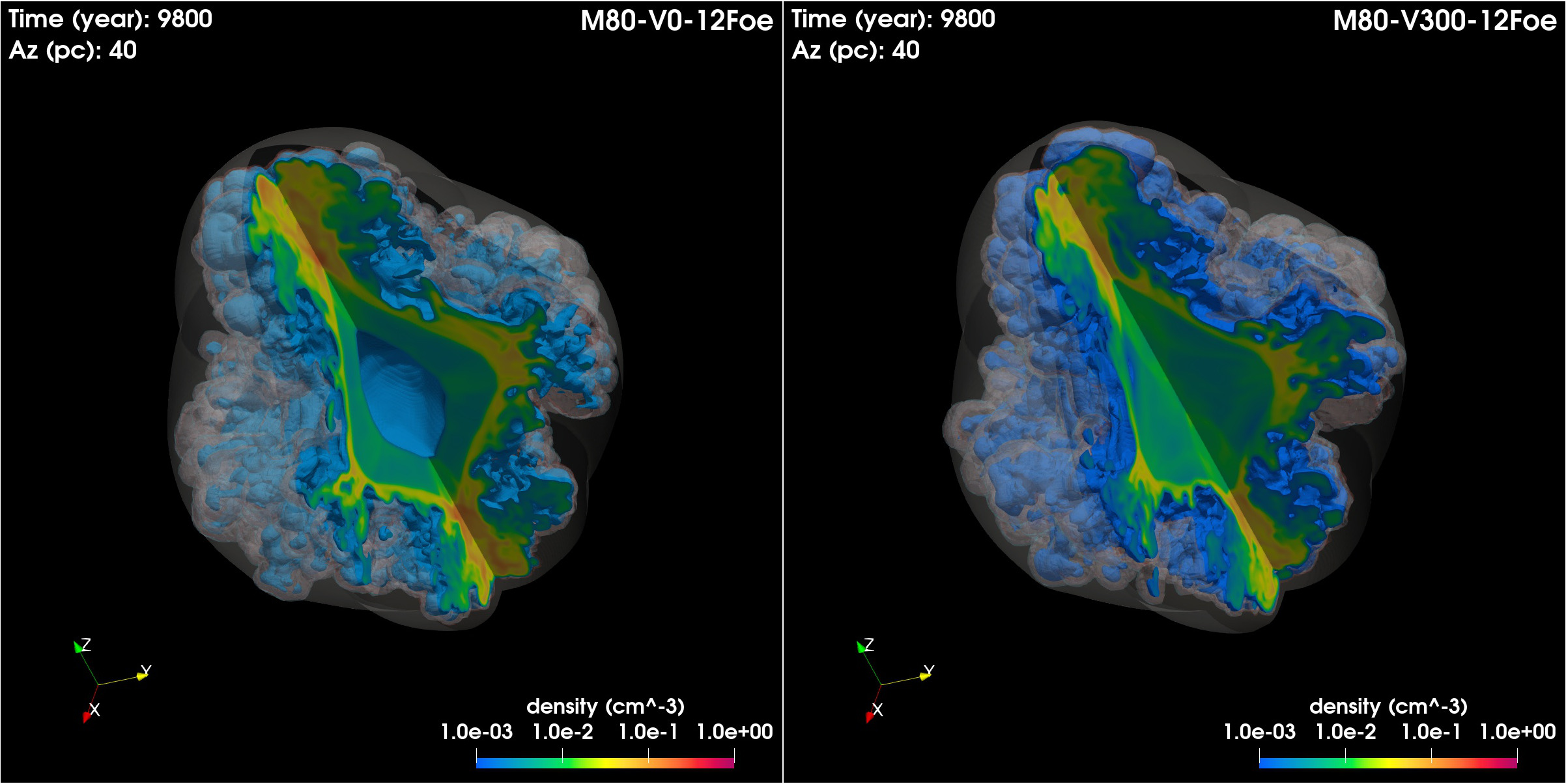}
    \caption{Density distributions for the ejecta of the models 
    M80-V0-12Foe (left panel) and M80-V300-12Foe (right panel)
    after $\sim 10000$ years of evolution.
    The isosurfaces and the orientation of the system are defined as in Fig.~\ref{Fig:hydro}.}
    \label{Fig:hydro_comp}
  \end{figure*}
  
  The evolution and density distribution of the SNRs in models with the same explosion energy but different masses of the progenitor stars (models M60-V0-1Foe and M80-V0-1Foe) are very similar.   
  The reason for this resemblance is that the increase of mass in 
  the progenitor is not reflected in the mass of the ejecta, as all 
  the additional mass may have been previously expelled into the CSM 
  or lost in the fallback immediately after the SN event (see second and third 
  columns in Table~\ref{Tab:domain}). This is also true for 
  models with low explosion energy and a non-zero rotation velocity (i.e., M60-V300-1Foe and M80-V300-1Foe). Thus, the explosion energy plays a crucial role in determining the morphology of the remnant during the interaction with the inhomogeneous CSM.
  
  This is confirmed by the models with a high explosion energy. In fact, we found that the remnant evolution and its morphology change appreciably from models M60-V0-9Foe and M60-V300-9Foe to models M80-V0-12Foe and M80-V300-12Foe due to a higher explosion energy in the latter
  (see Table~\ref{Tab:domain}). As a consequence, the ejecta expand faster in models M80-V0-12Foe and M80-V300-12Foe than in the rest of the models and, thus, the reverse shock refocuses at later times. For instance, in model M80-V0-12Foe the internal ejecta have not been shocked by the reverse shock 
  at the end of the simulation (see left panel in Fig.~\ref{Fig:hydro_comp}), whereas the reverse shock refocused at the center of the explosion at t$\approx 6500$~years in model M60-V0-9Foe. In model M80-V300-12Foe, the ejecta mass is only slightly higher than in model M60-V0-9Foe (see Table~\ref{Tab:domain}), and the reverse shock refocuses at t$\approx 9000$~years (instead of t$\approx 6500$~years as in M60-V0-9Foe).
   
\begin{figure}
    \includegraphics*[width=\hsize]{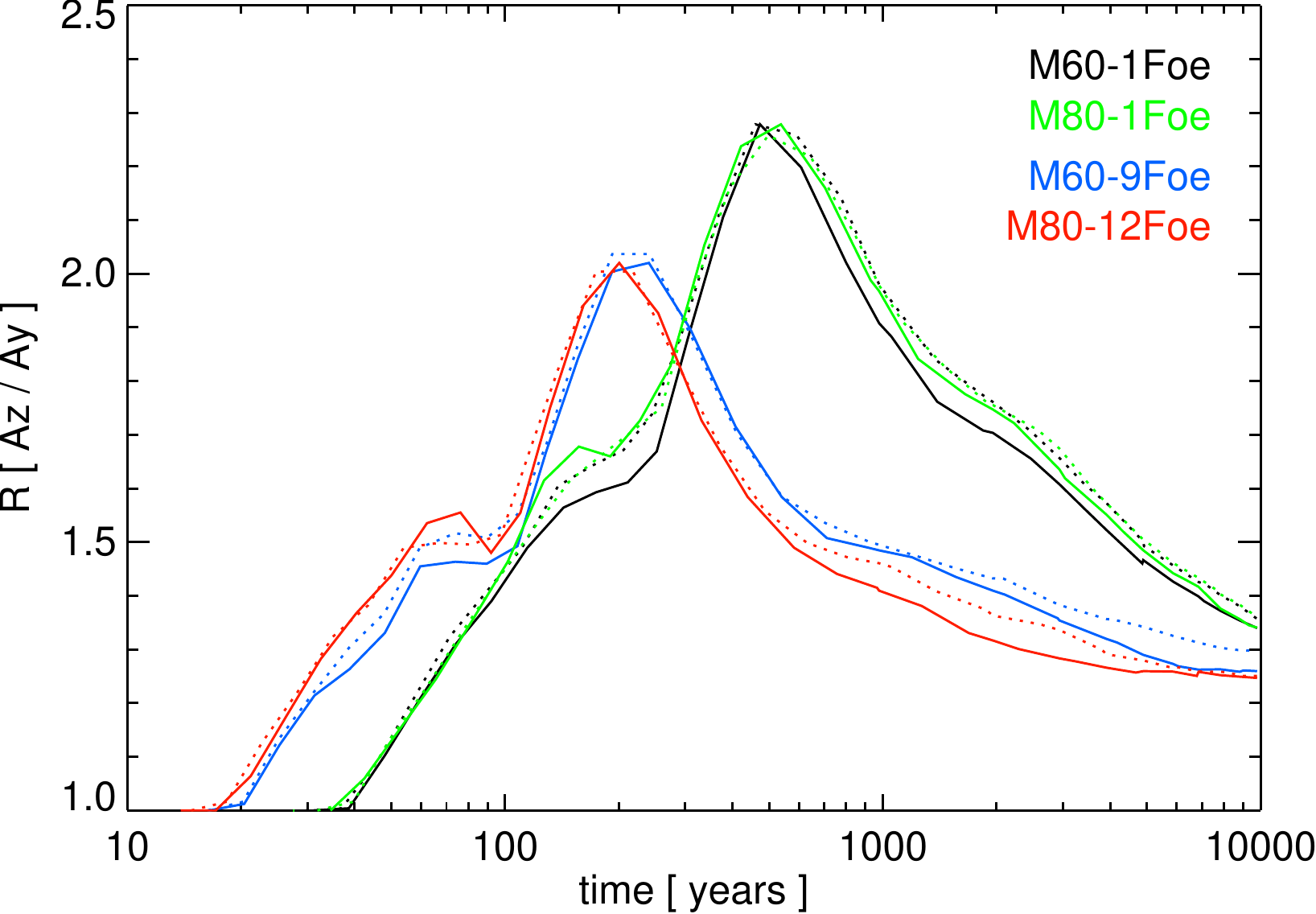}
    \caption{Degree of asymmetry of the remnant morphology, $R$, 
    versus time for the models
    explored (see Table~\ref{Tab:models}): M60-1Foe (black), 
    M80-1Foe (green), M60-9Foe (blue), M80-12Foe (red). 
    In each case, solid lines represent models with $V_{\rm rot}=0$, and 
    dotted lines the ones with $V_{\rm rot}=300$~km~s$^{-1}$.}
    \label{Fig:asym}
  \end{figure}

  Our simulations show that the remnant morphology is characterized by a large scale asymmetry due to the interaction with the dense shells, which changes significantly during the evolution. We can describe the degree of asymmetry of the remnant by the parameter $R = A_{\rm z }/A_{\rm xy}$, where $A_{\rm z}$ and $A_{\rm xy}$ are the distances of the forward shock from the center of the explosion along the $z$-axis and in the $xy$ plane, respectively. In Figure~\ref{Fig:asym}, we plot the parameter $R$ versus the time of evolution (in logarithmic scale) for the different models explored (see Table~\ref{Tab:models}). The figure shows that the models present three well-differentiated phases: 1) the initial free expansion through the stellar wind; 2) the interaction with the two dense shells; 3) the later expansion through the wind. During the first phase, $R\approx 1$, which means that the remnant is almost spherically symmetric: the homologous expansion of the remnant preserves the symmetry of the ejecta in all the models explored. In the subsequent phase, when the ejecta start to interact with the two dense shells (see Fig.~\ref{Fig:initial}), the remnant becomes progressively 
  more asymmetric, reaching the maximum asymmetry after the interaction with the outer shell. In fact, when the forward shock starts to interact with the inner dense shell, the $z$-axis becomes the preferred direction of the remnant expansion and $R$ increases. The growth of $R$ stops shortly after the interaction with the inner shell, and then continues growing while interacting with the outer shell. 
  In Fig.~\ref{Fig:asym} we identify two groups of models: those 
  with low and those with high explosion energy.
  Models with high explosion energy start to interact earlier with the dense shells and reach a maximum degree of asymmetry ($\sim 2$) at $t\approx 240$~yr (see Fig.~\ref{Fig:asym}; see also top right panel in Fig.~\ref{Fig:hydro_en}); models with low explosion energy start to interact later with the dense shells and reach a maximum degree of asymmetry ($\sim 2.3$) at $t\approx 500$~yr (see Fig.~\ref{Fig:asym}; see also top right panel in Fig.~\ref{Fig:hydro}). 
  Models with different initial mass of the progenitor star but with the same explosion energy follow an analogous evolution. In fact, our simulations show that models with the same explosion energy lead to very similar ejecta masses despite the main-sequence mass of the progenitor star being either 60 or $80\,M_{\odot}$ (see Table~\ref{Tab:domain}). Since the ejecta mass (and not the main-sequence mass of the progenitor star) is relevant in the evolution of the remnant, for the cases explored here, the explosion energy turns out to be the most relevant factor in the dynamical evolution of the SNR.
  We also note that models with higher explosion energy evolve in shorter timescales and reach a lower degree of asymmetry compared to those with lower explosion energy. In fact, the timescale of evolution reflects the expansion velocity of the remnant, which is higher in models with higher explosion energy. On the other hand, the degree of asymmetry of the remnant depends on both the energy of the explosion and the density contrast of the shells. The changes in the degree of asymmetry of the remnant are less evident than the changes in the timescale because, for the asymmetry, a central role is played by the density contrast of the shells, which is the same in all the models considered here.
  In any case, models with the same other parameters and either with or without rotation follow similar evolution. In the last phase, the ejecta continue to expand through the ambient medium in all directions while becoming again more symmetric. In this phase, the parameter $R$ gradually decreases showing the progressive reduction of the asymmetry. 
  In all the cases, the large-scale morphology of the SNR keeps memory of the early interaction of the remnant with the inhomogeneous CSM for almost 10000 years (namely the period covered by our simulations).
  
\subsection{Mass distribution in velocity space}
\label{Sec:mass}

\begin{figure*}
    \includegraphics*[width=\hsize]{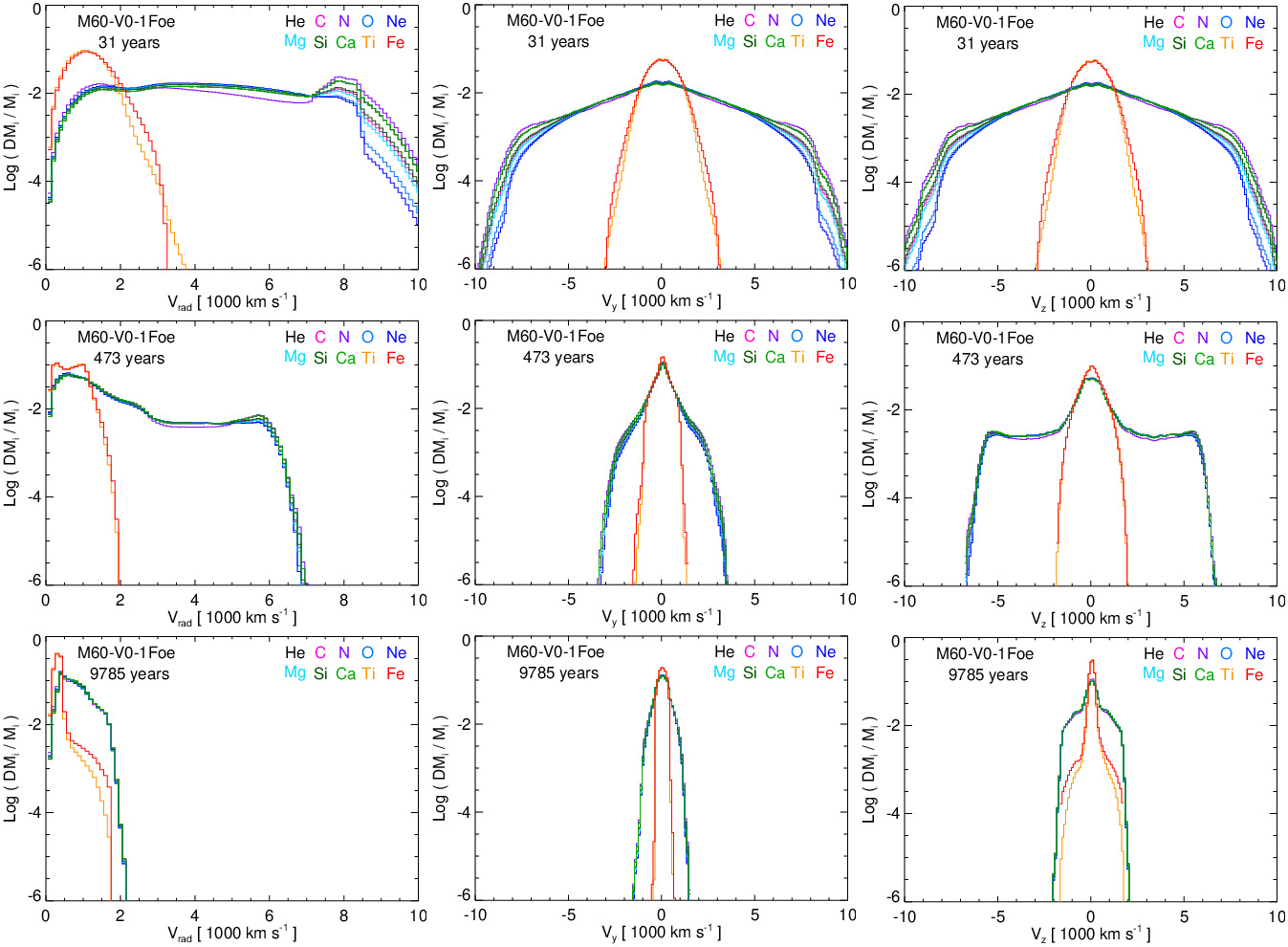}
    \caption{Mass distributions of $^4$He, $^{12}$C, $^{14}$N, $^{16}$O, 
    $^{20}$Ne, $^{24}$Mg, $^{28}$Si, $^{40}$Ca, $^{44}$Ti, and $^{56}$Ni 
    versus the radial velocity, V$_{\mathrm{rad}}$ (first column), 
    the velocity along the $y$-axis, V$_{\mathrm{y}}$ (second column), 
    or the $z$-axis, V$_{\mathrm{z}}$ (third column), for model M60-V0-1Foe 
    at different evolution times (increasing from top to bottom).
    Top row: just before the interaction with the innermost shell;
    middle row: at the maximum degree of asymmetry after the interaction 
    with the two shells; bottom row: near the end of the evolution.}
    \label{Fig:mass_low}
  \end{figure*}

\begin{figure*}
    \includegraphics*[width=\hsize]{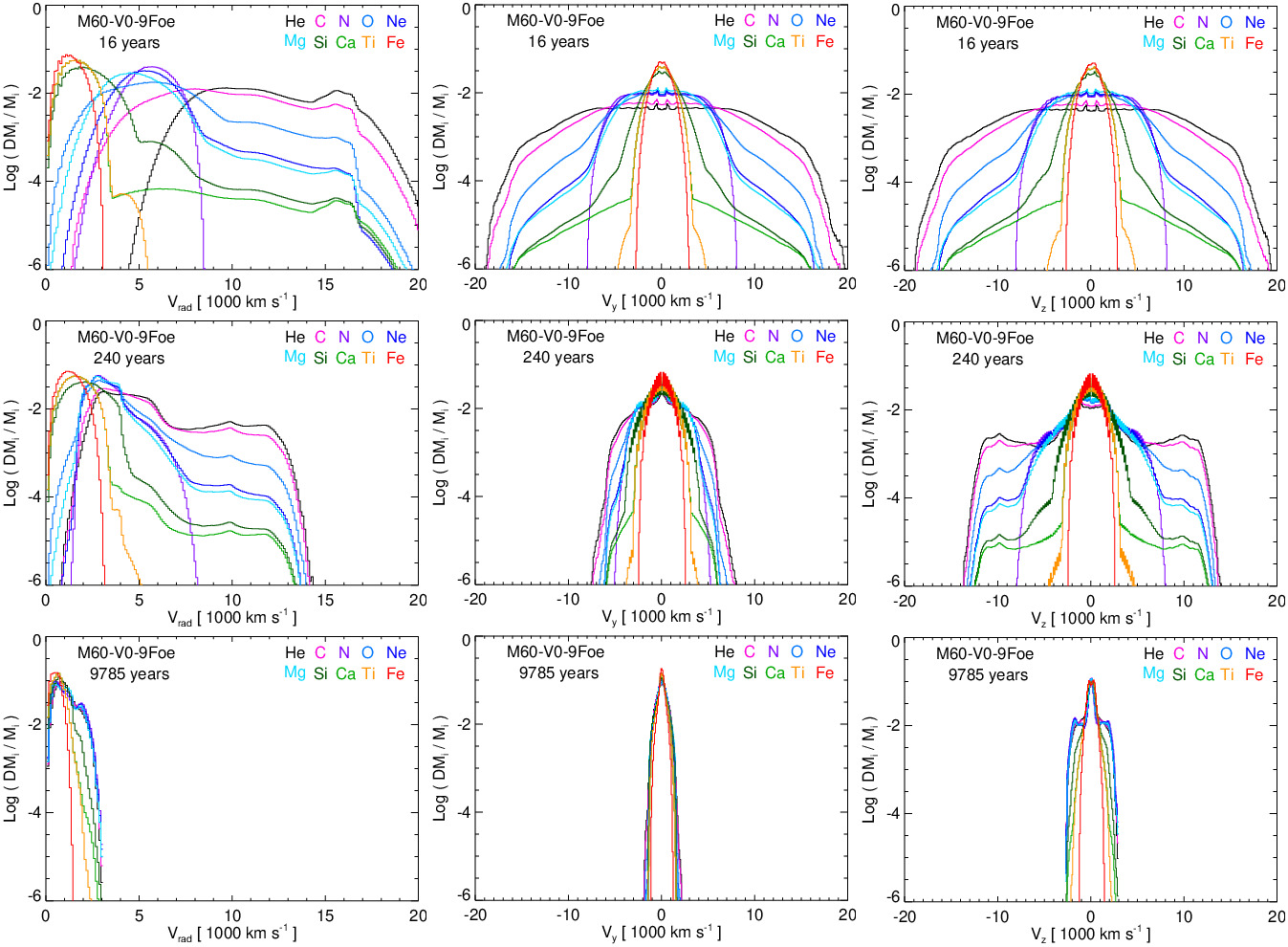}
    \caption{Same as in Fig.~\ref{Fig:mass_low} but for model M60-V0-9Foe.}
    \label{Fig:mass_high}
  \end{figure*}

  HD instabilities that develop during the interaction of the ejecta with 
  the reverse shock determine the mixing of the shock-heated ejecta in 
  the region between the forward and the reverse shocks, and thus the 
  distribution of the chemical elements at different evolution times of
  the SNR. The mixing between layers of different chemical composition
  during the evolution is reflected in the velocity distributions at 
  different ages of the SNR.
  Figures~\ref{Fig:mass_low}~and~\ref{Fig:mass_high} show the mass 
  distributions of selected elements, in models M60-V0-1Foe and 
  M60-V0-9Foe respectively, versus the radial velocity, V$_{\mathrm{rad}}$
  (first column), and the velocity along the LoS when the point of view lies either on the $y$-axis, V$_{\mathrm{y}}$ (second column), or on the $z$-axis, 
  V$_{\mathrm{z}}$ (third column), at different evolution times 
  (increasing from top to bottom). In all the figures presented, 
  $DM_{\mathrm{i}}$ is the mass of the $i$th element in the velocity 
  range $[v,v+dv]$, where $dv=100$~km~s$^{-1}$ is the velocity binning,
  and $M_{\mathrm{i}}$ is the total mass of the $i$th element.
  In the upper row, we present the distributions after the homologous 
  expansion of the SNR through the stellar wind and immediately before 
  the interaction with the innermost toroidal shell; in the middle row, 
  we show the distributions after the interaction with the two shells 
  when the SNR reaches its maximum degree of asymmetry (see 
  Fig.~\ref{Fig:asym}); finally, in the lower row, we plot the 
  distributions at the end of the simulation.
  
  Before the interaction with the two toroidal shells, the mass 
  distributions are similar to those of the initial condition,
  and the distributions versus V$_{\mathrm{y}}$ and V$_{\mathrm{z}}$ 
  are almost the same (see upper row in 
  Figs.~\ref{Fig:mass_low}~and~\ref{Fig:mass_high}). This is a result
  of the spherically symmetric initial conditions considered and the homologous 
  expansion of the SNR through the stellar wind. When the remnant 
  starts to interact with the dense shells, the mass distributions of the species vs. the LoS velocity is different if the point of view lies on the $y$ or on the $z$ axis. In the first case, the species can reach a maximum velocity which is significantly smaller than in the second case, due to the slowdown of the ejecta caused by the interaction with the shells. If the point of view lies on the $z$ axis, the ejecta appear to expand faster, being the $z$-axis the preferential direction of expansion. These differences between
  the mass distributions versus V$_{\mathrm{y}}$ and V$_{\mathrm{z}}$
  are maintained also after the interaction with the dense shells 
  (see Figs.~\ref{Fig:mass_low}~and~\ref{Fig:mass_high}), although they reduce, following the decrease in the degree of asymmetry of the remnant (see Fig.~\ref{Fig:asym}). We note that the mass distributions vs the LoS velocities V$_{\mathrm{y}}$ and V$_{\mathrm{z}}$ are symmetric with respect to the zero velocity. This is due to the idealized density distribution of the CSM (in particular the two nested shells) adopted here, which is characterized by cylindrical symmetry about the $z$-axis (see Fig.~\ref{Fig:initial} and Sect.\ref{Sec:num}).
  An asymmetry in the density distribution of the shells would be reflected in asymmetric mass distributions vs. the LoS velocities V$_{\mathrm{y}}$ and V$_{\mathrm{z}}$. 
  
  In model M60-V0-1Foe most chemical elements follow similar 
  shapes and are distributed in a broad maximum below 
  $\approx 10000$~km~s$^{-1}$ after 31 years of evolution
  (see upper row in Fig.~\ref{Fig:mass_low}), 
  immediately before the interaction with the innermost dense shell 
  (see left top panel in Fig.~\ref{Fig:hydro}).
  The exceptions are the unstable $^{44}$Ti and $^{56}$Ni (and their decaying products $^{44}$Ca, $^{56}$Co and $^{56}$Fe at late evolution times) which follow a 
  narrower distribution with maximum below $\approx 4000$~km~s$^{-1}$. 
  However, we note that in all the models with low explosion energy the total 
  masses of $^{44}$Ti and $^{56}$Ni (and, therefore, their decaying products) are very low in comparison with 
  lighter elements (see Table~\ref{Tab:mass}), as heavy elements go 
  primarily to the fallback during the collapse of the star (\citealt{lim18}). 
  The distributions of intermediate-mass and light elements at velocities 
  larger than $\approx 7000$~km~s$^{-1}$ have similar shapes with a slope 
  much steeper than in the initial condition (soon after the shock 
  breakout). The steepening of their slopes is a sign that these ejecta have already passed through the reverse shock and their similar shape is a sign of efficient mixing in the region between 
  the reverse and forward shocks. 
  The ejecta continue their expansion slowed down as a consequence of
  their interaction with the dense CSM. After interacting with both 
  dense shells (see right top panel in Fig.~\ref{Fig:hydro}), the 
  external layers of the ejecta have a maximum radial velocity of 
  $\approx 7000$~km~s$^{-1}$ (see left middle panel in 
  Fig.~\ref{Fig:mass_low}). At this time the remnant reaches its maximum 
  degree of asymmetry (see Fig.~\ref{Fig:asym}) with its preferential 
  direction of expansion along the $z$-axis, being 
  V$_{\mathrm{y}}<4000$~km~s$^{-1}$ for all the species (see central 
  and right middle panels in Fig.~\ref{Fig:mass_low}). 
  At $t\approx 473$~yr, the reverse shock starts to interact with the
  internal layers of the ejecta with velocities lower than 
  $\approx 2000$~km~s$^{-1}$ and finally focuses on the $z$ axis at $t\approx 1000$~yr.
  Then, the ejecta continue to expand in all directions lowering their 
  degree of asymmetry (see Figs.~\ref{Fig:hydro},~\ref{Fig:vel}~and~
  \ref{Fig:asym}). At the end of the simulation 
  V$_{\mathrm{rad}}<2000$~km~s$^{-1}$, V$_{\mathrm{y}}<1500$~km~s$^{-1}$
  and V$_{\mathrm{z}}<2000$~km~s$^{-1}$.
  
  In model M60-V0-9Foe the mass distributions versus the velocity
  are less homogeneous than in models with low explosion energy. Most light and
  intermediate-mass elements are distributed in a broad maximum below 
  $\approx 20000$~km~s$^{-1}$ after 16 years of evolution
  (see upper row in Fig.~\ref{Fig:mass_high}), 
  immediately before the interaction with the innermost dense shell 
  (see left top panel in Fig.~\ref{Fig:hydro_en}).
  The $^{14}$N, $^{44}$Ti and $^{56}$Ni instead follow 
  a narrower distribution; in particular, for $^{44}$Ti and $^{56}$Ni (and their decaying products)
  which are present mainly in the internal layers of ejecta, the
  maximum velocity of propagation is $\approx 5000$ and $\approx 3000$
  km~s$^{-1}$, respectively (see upper row in Fig.~\ref{Fig:mass_high}).
  In this case, the similarities shown by most of the distributions of 
  intermediate-mass and light elements at velocities larger than 
  $\approx 16000$~km~s$^{-1}$ (see upper row in 
  Fig.~\ref{Fig:mass_high}), indicate an efficient 
  mixing in the region between the reverse and forward shocks. 
  Similarly to the low explosion energy case, the expansion of the ejecta 
  in the equatorial plane is slowed down during their interaction with 
  the dense CSM, reaching the remnant its maximum degree of asymmetry 
  at $t \approx 240$~yr (see Fig.~\ref{Fig:asym} and right top panel 
  in Fig.~\ref{Fig:hydro_en}). At this time, V$_{\mathrm{rad}}<15000$~km~s$^{-1}$,
  V$_{\mathrm{y}}<8000$~km~s$^{-1}$, and V$_{\mathrm{z}}<15000$~km~s$^{-1}$ (see middle row 
  in Fig.~\ref{Fig:mass_high}). The reverse shock traveling through
  the ejecta has not yet reached the internal layers, dominated 
  by $^{44}$Ti and $^{56}$Ni and their decaying products (see middle row in Fig.~\ref{Fig:mass_high}).
  After this interaction phase, the ejecta continue to expand in all 
  directions lowering their degree of asymmetry (see 
  Figs.~\ref{Fig:hydro_en},~\ref{Fig:vel_en}~and~\ref{Fig:asym}). 
  At the end of the simulation V$_{\mathrm{rad}}<3000$~km~s$^{-1}$, 
  V$_{\mathrm{y}}<2000$~km~s$^{-1}$ and 
  V$_{\mathrm{z}}<3000$~km~s$^{-1}$. We find velocities slightly 
  higher than in the low explosion energy case at t$\approx10000$~yr, but 
  in model M60-V0-9Foe $Az=38$~pc (in M60-V0-1Foe, $Az=24$~pc at the same time) and the remnant has already reached
  a low degree of asymmetry (see Fig.~\ref{Fig:asym} and bottom
  right panels in Figs.~\ref{Fig:hydro}~and~\ref{Fig:hydro_en}).
  
\begin{figure*}
    \includegraphics*[width=\hsize]{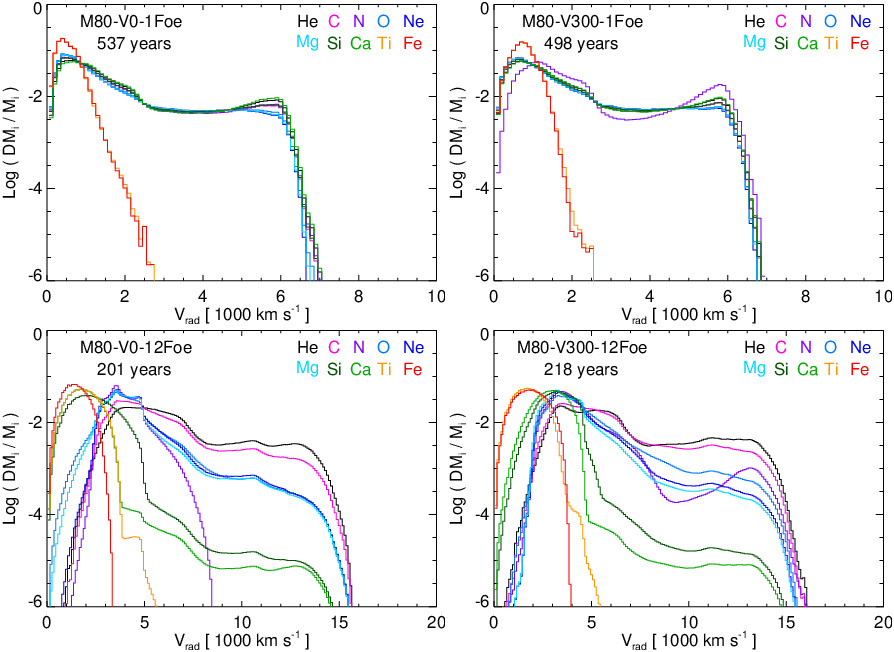}
    \caption{Mass distributions of $^4$He, $^{12}$C, $^{14}$N, $^{16}$O, 
    $^{20}$Ne, $^{24}$Mg, $^{28}$Si, $^{40}$Ca, $^{44}$Ti, and $^{56}$Ni 
    versus radial velocity, $V_\mathrm{rad}$, for models M80 at the
    time they reach their maximum degree of asymmetry. 
    The figure shows models either with low (upper panels) or with high 
    (lower panels) explosion energy, and models either without (on the left) 
    or with (on the right) initial rotation velocity of the star.}
    \label{Fig:mass_rot}
  \end{figure*}

  The mass distributions for the rest of models explored with low and high explosion energy are very similar to those presented for models
  M60-V0-1Foe and M60-V0-9Foe, respectively.
  In Fig.\ref{Fig:mass_rot}, we plot the mass distributions versus 
  $V_\mathrm{rad}$ for models M80 after the interaction with the two 
  toroidal shells, at the time they reach their maximum degree of asymmetry
  (see Fig.\ref{Fig:asym}).
  Models M80-V0-1Foe and M80-V0-12Foe presented in left panels in 
  Fig.~\ref{Fig:mass_rot} show very similar distributions to those presented 
  for models M60-V0-1Foe and M60-V0-9Foe, respectively (see left middle 
  panels in Figs.~\ref{Fig:mass_low}~and~\ref{Fig:mass_high}.
  In the analogous models but with an initial rotation velocity of the star (see right panels in Fig.~\ref{Fig:mass_rot}), 
  the only difference we observe is the higher abundance of $^{14}$N present 
  in the He core due to the effect of the mechanical instabilities induced 
  by rotation \citep{lim18}.

\subsection{Spatial distribution and chemical composition of the ejecta}
\label{Sec:chemical}

  We explored the chemical evolution of the ejecta for
  the 10 different species considered in this work
  ($^4$He, $^{12}$C, $^{14}$N, $^{16}$O, $^{20}$Ne, $^{24}$Mg, $^{28}$Si, 
  $^{40}$Ca, $^{44}$Ti, $^{56}$Ni). 
  The most abundant species in all the models analyzed are the 
  $^{16}$O and the $^{12}$C (see Table~\ref{Tab:mass}). 
  The total mass of the ejecta is much lower than the mass of the fallback 
  in all the models with low explosion energy (see Table~\ref{Tab:domain}).
  Thus, in all these cases, the masses of the heaviest species, namely the 
  $^{44}$Ti and the $^{56}$Ni (and, therefore, their decaying products), are very low (see Table~\ref{Tab:mass}). In models with low explosion energy, therefore, we find quite 
  homogeneous distributions of light and intermediate mass elements in the 
  ejecta, with a very residual presence of heavier elements. 
  
\begin{figure*}
    \includegraphics*[width=\hsize]{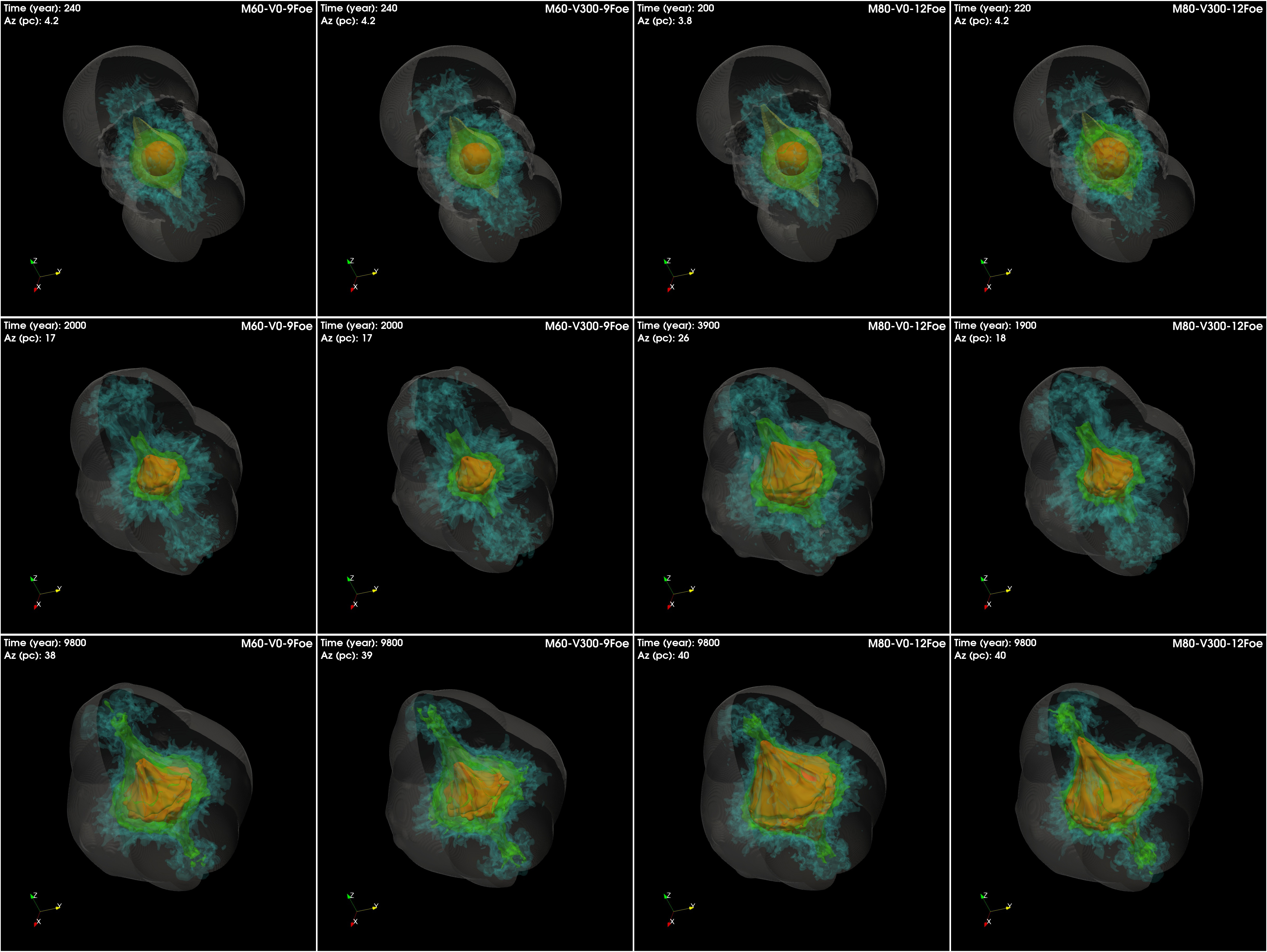}
    \caption{Density distributions for the ejecta rich in Fe (orange), 
    Si (green), and O (blue) for the high explosion energy models 
    at different evolution times (increasing from top to bottom).
    The opaque irregular isosurfaces correspond to a value of density 
    at 1\% of the peak density in each case.
    The semi-transparent surface marks the position of the forward shock; 
    the yellow semi-transparent surface barely visible in top panels represent the 
    position of the reverse shock. The orientation of the system is the same of G26 (see Fig.~\ref{Fig:initial}).}
    % Movie?
    \label{Fig:che}
  \end{figure*}

  In models with high explosion energy instead, we find a less homogeneous 
  distribution of chemical elements in the ejecta. In these cases, the mass of the 
  ejecta is much larger than the mass of the fallback (see 
  Table~\ref{Tab:domain}) and a significant amount of $^{44}$Ti and $^{56}$Ni (and their decaying products) is present
  in the internal layers (see Table~\ref{Tab:mass}).
  In Figure~\ref{Fig:che}, we show the density distribution for the ejecta 
  rich in Fe\footnote{At this evolution time, almost all the $^{56}$Ni has already 
  decayed in $^{56}$Fe.}, Si and O for the four high explosion energy models 
  explored at different evolution times (increasing from top to bottom).
  After the interaction with the two toroidal shells, at $t\approx 200$~yr
  (see upper panels in Fig.~\ref{Fig:che}), the four distributions show 
  similar characteristics: the O, distributed in the external layers 
  of the ejecta and already heated by the reverse shock, forms an 
  elongated structure along the $z$ axis; the Si (the spherical green surface) is 
  starting to interact with the reverse shock (the yellow transparent 
  surface visible in the upper panels); and the Fe (the orange spherical most 
  internal surface) is still unshocked.
  In model M80-V300-12Foe (see right top panel in Fig.~\ref{Fig:che})
  the total mass of Fe is higher (see Table~\ref{Tab:mass}) and the
  Si has already been partially shocked by the reverse shock.
  The forward and reverse shocks continue to propagate in opposite 
  directions and, after $\sim 2000$ years of evolution, the ejecta start
  to form a Si-rich jet-like structure along the $z$-axis in most of the models
  (see middle panels in Fig.~\ref{Fig:che}).
  In model M80-V0-12Foe this occurs later ($\sim 4000$~yr) due to the 
  higher mass of the ejecta and the slower propagation of the reverse 
  shock (see Sec.~\ref{Sec:hd}).
  At the end of the simulation, all the high explosion energy models show 
  Fe-rich internal ejecta distributions surrounded by an elongated 
  Si-rich structure with a more diffuse O-rich ejecta around 
  (see lower panels in Fig.~\ref{Fig:che}). 
  The Fe-rich internal structure is more extended and elongated 
  in models M80.
  
  The distribution of material in Fig.~\ref{Fig:che} that could be 
  observed can be very different depending on the LoS considered. 
  For instance, when the LoS lies on the equatorial plane (the $xy$ plane), 
  the ejecta projected in the plane of the sky form 
  a jet-like structure rich in Fe and Si, surrounded by a more diffuse O-rich material in the external layers.  
  When integrating along the $z$-axis instead, the projected ejecta form a 
  spherical central structure with mixed composition surrounded by 
  more diffuse O-rich ring. 
  Finally, in the case of the LoS of G26 (see Fig.~\ref{Fig:che}),
  we expect to observe a structure of projected ejecta intermediate between the two extreme cases described above.

%--------------------------------------------------------------------
\section{Summary and discussion}
\label{Sec:conclusions}

  We investigated what the SNR would be like, if the LBV candidate G26 exploded as SN. To this end, we modeled the evolution of a LBV exploding as a core-collapse SN, starting 
  immediately after the shock breakout. Then, we followed the transition 
  from the SN to the SNR phase and the interaction of the remnant with 
  the inhomogeneous pre-SN environment for $\approx 10000$~yr and for 
  eight different parent SNe from LBV-like stars compatible with G26 (see Sec.~\ref{Sec:num}). 
  The SNR models differ from each other as for the explosion energy and for the characteristics of the progenitor star 
  (see Table~\ref{Tab:models}), which have been selected from the pre-SN 
  models investigated by \cite{lim18} and prescribed as 
  initial conditions in this work. 
  The pre-SN environment is the same 
  for all the models and is based on infrared and radio observations of G26
  (see Fig.~\ref{Fig:initial}).
  
  Our models show three well differentiated phases in the evolution of the remnant:
  1) its expansion through the innermost CSM modeled here as a $r^{-2}$ stellar wind; 2) its interaction with the dense CSM characterized by two dense toroidal shells; and 3) the expansion of the remnant through an almost uniform ambient environment.
  After the homologous expansion through the stellar wind, at 
  $t\approx 18-40$~yr, the forward shock starts to interact with the 
  innermost dense toroidal shell characterizing the CSM (see Fig.~\ref{Fig:initial}). 
  The interaction determines a strong slowdown of the forward shock 
  in the equatorial plane, and a strengthening of the reverse 
  shock traveling through the ejecta (see Figs.~\ref{Fig:vel}~and~\ref{Fig:vel_en}). 
  During this phase the $z$-axis 
  becomes the preferred direction of expansion of the blast wave and the remnant progressively becomes more asymmetric, reaching its maximum degree of asymmetry 
  ($R$) after the interaction with the outermost dense shell (see 
  Fig.~\ref{Fig:asym}), at $t\approx 200-500$~yr. 
  Finally, the remnant continues to expand through the wind while $R$
  decreases. After $\approx 10000$~years of evolution the ejecta 
  have slowed down their expansion and they show an elongated shape 
  forming a broad jet-like structure with maximum density along the $z$-axis, 
  which extends for $\approx 24-38$~pc from the center of the explosion (see Figs.~\ref{Fig:hydro}~and~\ref{Fig:hydro_en}).
  These jet-like structures are the relic of the early interaction of the 
  remnant with the inhomogeneous CSM that was sculpted by violent mass-loss 
  events occurred in the latest phases of evolution of the progenitor star.
  
  Among the cases explored (see Table~\ref{Tab:models}), there are
  two classes of models: those with low ($10^{51}$~erg) and those with high 
  ($9-12\times 10^{51}$~erg) explosion energy (see Fig.~\ref{Fig:asym}). The initial
  mass of the progenitor and the explosion energy determine the mass,
  the expansion velocity and the chemical composition of the ejecta.
  Models with higher explosion energy evolve in shorter timescales 
  and reach a lower degree of asymmetry compared to those with low explosion energy. In fact, in the latter models, the remnant starts to interact later with the 
  dense shells but is more affected by the CSM (which limits the ejecta motion in the equatorial plane and triggers a 
  strengthening of the reverse shock traveling through the ejecta) 
  and reaches a higher degree of asymmetry (see Fig.~\ref{Fig:asym}). We note that the explosion energy influences the timescale of evolution more than the remnant asymmetry because the timescale reflects the expansion velocity of the remnant (that is higher for higher explosion energy), whilst the asymmetry depends on both the explosion energy and the density contrast of the shells, with the latter being the same in all the models. Low explosion energy models, all have a relatively low mass of ejecta ($< 4 M_{\odot}$) as most
  of the material has been previously expelled into the CSM or lost 
  in the fallback during the SN event (see Table~\ref{Tab:domain}). 
  As a consequence, the mass of the heavier species (namely
  $^{44}$Ti and $^{56}$Ni and their decaying products) is very low in all the low explosion energy models
  (see Table~\ref{Tab:mass}). The spatial distribution of the light 
  and intermediate-mass elements is quite uniform in all the low
  explosion energy models, being the $^{12}$C and $^{16}$O the most abundant
  from the species considered.
  All the models show an efficient mixing in the region between the
  reverse and forward shocks (see Fig.~\ref{Fig:mass_low}). 
  The only difference we observe between analogous models either with or without 
  rotation is the higher abundance of $^{14}$N present 
  in the He core due to the effect of mechanical instabilities 
  induced by rotation in models with V$_{\mathrm{rot}}=300$~km~s$^{-1}$ (see Fig.~\ref{Fig:mass_rot}; see also \citealt{lim18}).
  
  In the models with high explosion energy, the expansion of the ejecta
  in the $xy$-plane is also slowed down due to the interaction
  with the dense shells but this effect is less pronounced than in models with low explosion energy. As
  a consequence, the degree of asymmetry reached in these models is lower (see Fig.~\ref{Fig:asym}) and the
  reverse shock refocuses at later times (see Fig.~\ref{Fig:hydro_en}).
  The high explosion energy models show a higher mass of ejecta and a less 
  homogeneous chemical distribution of ejecta than the low explosion energy models.
  After $\approx 2000-4000$ years of evolution, the reverse shock
  starts to interact with the innermost layers of ejecta, forming
  a Si-rich jet-like structure along the $z$-axis. At the end of the 
  simulation ($t\approx 10000$~yr), all the high explosion energy models show 
  Fe-rich internal ejecta distributions surrounded by an elongated 
  Si-rich structure with a more diffuse O-rich ejecta 
  around (see lower panels in Fig.~\ref{Fig:che}). The Fe-rich 
  internal structure is more extended and elongated in models M80, namely those with $80\,M_{\odot}$ on the zero age main sequence. We stress here that these elongated features originate from the interaction of the remnant with the inhomogeneous ambient medium and do not reflect large-scale asymmetries left from the earliest phases of the SN explosion and, in some cases, developed from stochastic processes (convective overturn and the standing accretion shock instability) during the first seconds of the SN blast (e.g.,~\citealt{won15, won17}). 
  In all the cases examined, therefore, we found that the remnant morphology 
  keeps memory of the early interaction of the remnant with the inhomogeneous 
  CSM (the shells), even thousands of years after the SN. In other words, 
  the effects of mass-loss events occurred in the latest phases of the 
  progenitor star evolution could be still encoded in the asymmetries of 
  remnants up to 10000 years old (the time covered by our simulations).
  
  In this work, we studied the characteristics of remnants of 
  core-collapse SNe from LBV stars, which show a strong interaction with
  their CSM. Given the complex structure and high density of the CSM around LBVs, the characteristics of the CSM where a LBV goes SN play a fundamental role 
  in determining the properties of its remnant. In fact, fixing the structure of the CSM consistently with that inferred from observations of G26, we find a common morphology of the SNR for all the progenitors and explosion energies
  explored: elongated ejecta with an internal jet-like structure, 
  which is the result of the interaction with the highly inhomogeneous CSM. 
  
  It is worth to emphasize that we adopted, as a template, the LBV candidate G26 because the massive nebula in which the star is located has a structure and density distribution well characterized by observations \citep[see][]{uma12}. This has allowed us to define an idealized geometry for the
  dense CSM of G26 (see Fig.~\ref{Fig:initial}), consisting of two axially symmetric shells, each with $\sim 10\,M_{\odot}$. As expected, the two dense shells play a central role in modifying the expansion 
  of the forward shock and in driving a reflected shock through the ejecta. However, the CSM structure adopted here is specific of G26, but a wide range of masses and different geometries may characterize LBVs. In the light of our simulations, we can figure out what to expect if the mass of the shells or the geometry of the CSM are different from those of G26. If we consider the same geometry adopted here but more massive shells, we expect that the forward shock would be more 
  slowed down by the interaction with the denser shells, a stronger reflected 
  shock would be driven through the internal ejecta, and the remnant would have a morphology similar to that found here but with a higher degree of asymmetry.
  Conversely, if the remnant interacts with less massive shells, the forward 
  shock would expand fastly through the low density shells, a fainter 
  reflected shock would be driven backward through the ejecta and the remnant morphology would have a lower degree of asymmetry. As for the geometry of the CSM, we found that the two axially symmetric shells characterizing the CSM of G26 lead to a broad jet-like structure in the morphology of the remnant. Thus, we expect that large-scale inhomogeneities in the pre-SN CSM, possibly different from that adopted here, may be reflected in the final morphology and large-scale asymmetries of the remnant.
  
  We expect SNRs from LBV progenitors not to be common, since LBVs are considered to be brief transitional phase in the evolution of the quite rare most massive stars. Nevertheless, there are a few examples of remnants in the literature showing morphologies analogous to those modeled here: W50 \citep{dub98}; SNR G309.2-00.6 \citep{gae98};
  SNR~W44 \citep{she04}; SNR~S~147 \citep{dre05,gva06}. In all these cases, the remnants appear elongated along a preferential direction of expansion and, in some cases, they present some hints for a jet-like structure. The morphology characterized by ``ears'' of these SNRs has been interpreted by some authors as the result of jet-driven core-collapse SN mechanisms \citep[see][]{gri17}. If that were the case, the remnant could be inflated by jets that are launched during the explosion. Here, we have shown that similar morphologies can be also reproduced by the interaction of the remnant with a highly inhomogeneous and dense CSM as that surrounding LBVs, thus that could be the case for some of the remnants cited above. The role of the CSM in shaping the remnant morphology have been also recently investigated by \cite{chi21}.
  
  LBVs as explosive transients constitute a puzzling and still 
  poorly understood category. According to our models, a distinctive 
  property of SNe from LBV progenitors could be the high fallback of 
  matter soon after the core-collapse, specially for those with low 
  energy of the explosion. This implies the formation of a black hole 
  instead of a neutron star. In this case, we do not expect that the 
  SNRs from LBV progenitors normally host a detectable compact object 
  (a neutron star) in its interior. 
  LBVs are also extremely interesting SN progenitors due to the strong 
  interaction between their ejecta and the pre-existing slower and dense CSM.
  In this work, the pre-existing CSM has been described following the two-shell 
  environment identified in G26 by \cite{uma12} which, in turn, is a 
  simplified version of the CSM expected in a LBV exploding star. 
  Considering a more realistic description of the star wind and the CSM 
  close to the progenitor in future models could account for the high luminosities
  observed during the early phases of Type IIn SNe \citep[see][]{smi17_hsn}.
  
  The interaction of the blast wave with the dense shells can have also important 
  consequences for the acceleration of cosmic rays (CRs). During the interaction, 
  the remnant can become a strong $\gamma$-ray source and may provide evidence 
  for hadronic CRs acceleration \citep{byk18}. Furthermore, these remnants 
  can be factories of very energetic particles possibly up to PeV energies. 
  These very energetic particles are observed in the galactic CRs spectrum 
  but are not inferred from observations of SNRs. Since the blast wave from 
  the SNe of LBVs is very energetic during the first decades of the remnant 
  evolution, it could be possible that PeV particles are produced in the 
  interaction of the remnant with the dense shells \citep[e.g.][]{zir16}. 
  In this case, again we expect $\gamma$-rays that could be detected with 
  current and future instruments (e.g., the Cherenkov Telescope Array). 
  Self-consistent models as those analyzed here could allow to disentangle 
  the effects of interaction of the remnant with an inhomogeneous CSM from 
  those of the structure of the progenitor star. This could shed some light 
  into the last phases of evolution of massive stars and into the still thin 
  link between LBVs and Type IIn/IIb SNe.

\begin{acknowledgements}

   We thank the referee for useful comments and suggestions that 
   allowed us to improve the manuscript.
   We acknowledge the computing centre  of Cineca and INAF, under 
   the coordination of the "Accordo Quadro MoU per lo svolgimento di 
   attività congiunta di ricerca Nuove frontiere in Astrofisica: HPC 
   e Data Exploration di nuova generazione", and the HPC facility 
   (SCAN) of the INAF – Osservatorio Astronomico di Palermo for the 
   availability of computing resources and support.
   The PLUTO code, used in this work, was developed at the Turin 
   Astronomical Observatory in collaboration with the Department of 
   General Physics of Turin University and the SCAI Department of CINECA. 
   We acknowledge financial contribution by the INAF PRIN 2019 grant ``From massive stars to supernovae and supernova remnants: driving mass, energy and cosmic rays in our Galaxy'' and the INAF mainstream program ``Understanding Particle Acceleration in Galactic Sources in the CTA era''.

\end{acknowledgements}

%-------------------------------------------------------------------
%
\bibliographystyle{aa} % style aa.bst
\bibliography{biblio.bib} % my references biblio.bib
%
%-------------------------------------------------------------------
\begin{appendix}

\section{Online material}
\label{app:online}

\begin{itemize}
\setlength\itemsep{0.5em}
 
   \item \textbf{Movie~1}: Density distributions for the ejecta of the model M60-V0-1Foe. The opaque irregular isosurfaces correspond to a value of 
   density which is at 1\% of the peak density with one quadrant cut in 
   order to see the radial distribution. The semi-transparent surface marks 
   the position of the forward shock; the initially toroidal semi-transparent structures in red and cyan colors represent the inner and outer shells 
   in the CSM respectively. The system is oriented as G26, corresponding to 
   the rotation angles $i_x=30$º, $i_y=30$º, $i_z=25$º about the $x$, $y$, 
   and $z$ axes, respectively.
    
   \item \textbf{Movie~2}: Same as in Movie~1 but for the colors giving 
   the radial velocity in units of 1000~km~s$^{-1}$ on the isosurface.

   \item \textbf{Movie~3}: Same as in Movie~1 but for model M60-V0-9Foe. 
   The complete temporal evolution is available as online movie (Movie~3).

   \item \textbf{Movie~4}: Same as in Movie~3 but for the colors giving 
   the radial velocity in units of 1000~km~s$^{-1}$ on the isosurface.

\end{itemize}

\end{appendix}

\end{document}